\documentclass[pre,preprint,showpacs,showkeys,preprintnumbers]{revtex4}

\usepackage{graphicx}
\usepackage{dcolumn}
\usepackage{bm}
\usepackage{amsmath}
\usepackage[dvips]{color}

\begin{document}

\preprint{Aubry et al.}

\title[]{Detection and imaging in a random medium: a matrix method to overcome multiple scattering and aberration}

\author{Alexandre Aubry}
\affiliation{Institut Langevin, ESPCI ParisTech,\\
CNRS UMR 7587, Universit\'e Denis Diderot (Paris VII),\\
Laboratoire Ondes et Acoustique, 10 rue Vauquelin, 75005 Paris, France}

\author{Arnaud Derode}
\affiliation{Institut Langevin, ESPCI ParisTech,\\
CNRS UMR 7587, Universit\'e Denis Diderot (Paris VII),\\
Laboratoire Ondes et Acoustique, 10 rue Vauquelin, 75005 Paris, France}

\date{\today}

\begin{abstract}
We present an imaging technique particularly suited to the detection of a target embedded in
a strongly scattering medium. Classical imaging techniques based on the Born approximation fail in this kind of configuration because of multiply scattered echoes and aberration distortions. The experimental set up we consider uses an array of programmable transmitters/receivers. A target is placed behind a scattering medium. The impulse responses between all array elements are measured and form a matrix. The core of the method is to separate the single-scattered echo of the target from the multiple scattering background. This is possible because of a deterministic coherence along the antidiagonals of the array response matrix, which is typical of single scattering. Once this operation is performed, target detection is achieved by applying the DORT method (French acronym for decomposition of the time reversal operator). Experimental results are presented in the case of wide-band ultrasonic waves around 3 MHz. A 125-element array is placed in front of a collection of randomly distributed steel rods (diameter 0.8mm). The slab thickness is three times the scattering mean free path. The target is a larger steel cylinder (diameter 15 mm) that we try to detect and localize. The quality of detection is assessed theoretically based on random matrix theory and is shown to be significantly better than what is obtained with classical imaging methods.
Aside from multiple scattering, the technique is also shown to reduce the aberrations induced by an heterogeneous layer. 
\end{abstract}



\maketitle

\copyright{Copyright 2009 American Institute of Physics. This article may be downloaded for personal use only. Any other use requires prior permission of the author and the American Institute of Physics. The following article appeared in J. Appl. Phys. \textbf{106}, 044903 (2009), and may be found at \href{http://link.aip.org/link/?JAPIAU/106/044903/1}{http://link.aip.org/link/?JAPIAU/106/044903/1}}

\section{\label{sec:intro}Introduction}
Classical reflection imaging methods, such as echography or radar, are based on the same principle. One or several transducer(s) emit(s) a wave toward the medium to be imaged. The incident wave is reflected by the heterogeneities and the backscattered wave field is measured by the same sensor(s). The backscattered wave contains two contributions:
\begin{itemize}
\item A single scattering contribution (path $s$ in Fig.\ref{fig:setup}): the incident wave undergoes only one scattering event before coming back to the sensor(s). This is the contribution which is taken advantage of, because there is a direct relation between the arrival time
$t$ of the echo and the distance $d$ between the sensor and the scatterer, $t=2d/c$ ($c$ is the sound velocity). Hence an image of the medium's reflectivity can be built from the measured signals.
\item A multiple scattering contribution (path $m$ in Fig.\ref{fig:setup}): the wave undergoes several scattering events before reaching the sensor. Multiple scattering is expected to take place when scatterers are strong and/or concentrated. In this case there is no more equivalence between the arrival time $t$ and the depth of a scatterer. Thus, classical imaging fails when multiple scattering dominates.
\end{itemize}

To image an heterogeneous medium, one tries to reduce the influence of multiple scattering.
In that respect, multiple sensors arrays are a great improvement, since coherent beamforming can be achieved at emission and reception \cite{angelsen}. It consists in focusing the transmitted wave at the desired point by applying the appropriate time delays to each array element. In the reception mode, the same delays are applied to the received signals before they are summed. Single scattering signals coming from a target located at the focus add up coherently, whereas the summation is expected to be incoherent for multiple scattering signals arriving at the same time. The gain in single-to-multiple scattering provided by beamforming is proportional to the number of elements on the array. In medical imaging where multiple scattering is usually weak at standard ultrasonic frequencies, this operation is generally sufficient to correctly image the medium. But in other situations, multiple scattering can be so high that coherent beamforming fails. The resulting echographic image is pure speckle, with no direct connection with the medium's reflectivity. There can be false alarms that one can wrongly attribute to the presence of a strong reflector in the medium. Furthermore, aberration effects distort the wave front of the focused beam, which may generate secondary lobes or a displacement of the focal spot.

Our aim is to detect and image an echogene target embedded in a scattering medium. This issue has received considerable attention in the last decade \cite{chan,zhang,zhang2,chan2,borcea2,borcea,ishimaru,bal1,bal2,garnier}. As mentioned previously, classical imaging techniques may fail in such media because of multiple scattering and aberration effects. To solve this problem, various coherent interferometric imaging techniques have been suggested \cite{chan,zhang,zhang2,chan2,borcea2,borcea,garnier}. 
Nevertheless, they are shown to fail when the target is typically buried beneath one transport mean free path $l^*$ of the scattering medium \cite{borcea}. Another route towards target detection in highly scattering media is to tackle with the radiative transfer equation \cite{bal1,bal2}. However, this approach needs heavy numerical computations. Moreover, the final resolution of the image is poor since it is limited by $l^*$ instead of half the wave length $\lambda/2$.
This paper proposes an original approach to drastically reduce the multiple scattering contribution, which can hide the echo from targets (\textit{e.g.}, landmines, ducts, defects…) embedded in the earth \cite{norville2}, in concrete structures \cite{sornette,kozlov} (flaws, defects,...) or austenic steels \cite{bordier} for non destructive evaluation. Reducing the influence of multiple scattering is also a challenge in optical coherence tomography (OCT) \cite{karamata,karamata2,yadlowsky}, in seismology \cite{martini,shapiro}, in ultrasound imaging \cite{nelson,hedrick} or in radar \cite{tabbara}. In optics, correlation techniques have also been proposed to reduce the multiple scattering influence in dynamic light scattering experiments \cite{phillies,schatzel,meyer,zakharov,scheffold}. However, these methods only address the suppression of the multiple scattering contribution in the autocorrelation function of intensity. On the contrary, the approach we propose here can be dedicated to a much wider range of applications, since it basically applies to the wave field.

In this paper we will use ultrasonic waves in the MHz range for experimental demonstrations, but the technique can be applied to all fields of wave physics for which the multi-element array technology is available and provides time-resolved measurements of the amplitude and the phase of the wave field. The experimental situation is the following: the medium we want to image is placed in front of a multi-element
array (see Fig.\ref{fig:setup}). A pulse signal is sent from element $i$ and the wave backscattered by the medium is measured by element $j$. This operation is achieved for all possible transmitter/receiver couples. The set of $N^2$ responses forms a matrix $\mathbf{K}$ which constitues the global response of the medium. Unlike the multiple scattering contribution, single scattering signals exhibit a deterministic coherence along the antidiagonals of the array response matrix whatever the distribution of scatterers \cite{aubry09}. This particular feature can be taken advantage of to extract the single scattered waves, even though multiple scattering predominates. This ``single scattering filter''(SSF) yields a filtered matrix $\mathbf{K^F}$, ideally devoid of multiple scattering. 

Once the separation of single- and multiple-scattered waves is performed, the detection of the target is achieved by the DORT method \cite{prada,prada2} (French acronym for decomposition of the time reversal operator). It consists in a singular value decomposition (SVD) of the array response matrix. Actually the SVD is written $\mathbf{K} = \mathbf{U} \mathbf{\Lambda} \mathbf{V}^{\dag}$, where $\mathbf{\Lambda}$ is a diagonal matrix containing the real positive singular values $\lambda_i$ in a decreasing order ($\lambda_1>\lambda_2>...> \lambda_N$). $\mathbf{U}$ and $\mathbf{V}$  are unitary matrices whose columns are the normalized singular vectors $\mathbf{U_i}$ and $\mathbf{V_i}$. DORT has shown its efficiency in detecting and separating the responses of several scatterers in homogeneous or weakly heterogeneous media \cite{prada2}. Indeed, under the single scattering approximation and for point-like scatterers \cite{prada4,minonzio}, each scatterer is associated mainly with one significant eigenstate linked to a non zero singular value $\lambda_i$. The corresponding singular vector $\mathbf{V_i}$ is an invariant of the time reversal operator $\mathbf{K}\mathbf{K}^{\dag}$. Physically, each eigenvector of $\mathbf{K}\mathbf{K}^{\dag}$ (or singular vector of $\mathbf{K}$) corresponds to a wave that, when it is sent from the array, focuses onto the associated scatterer. Therefore, it is possible to focus selectively on the corresponding scatterer and obtain its image by backpropagating $\mathbf{V_i}$ either physically or numerically. 

However, in this study, the target is hidden behind a strongly scattering slab. If we apply the DORT method directly to the array response matrix $\mathbf{K} $, expecting that the target will be associated to the first singular value $\lambda_1$ and backpropagating numerically the corresponding singular vector $\mathbf{V_1}$, it fails because of multiple scattering. We will show that once the single- and multiple-scattering contributions have been separated, DORT can be applied to the filtered matrix $\mathbf{K^F}$, and successfully detects the target despite multiple scattering.
A detection criterion has to be applied to the first singular value $\lambda_1$ in order to decide if a target is detected or not. To that aim, we will refer to random matrix theory (RMT) \cite{tulino,sengupta} and to a recent work \cite{aubry} dealing with the statistical behavior of the matrix $\mathbf{K}$ in random media. The efficiency of the technique will also be evaluated from RMT and shown to be better than classical imaging techniques. Finally, the issue of aberration will be adressed. The SSF is shown to strongly diminish the aberration effects which occur in scattering media.

\section{\label{sec:exp_num}Experimental procedure}
\begin{figure}[htbp] 
\includegraphics{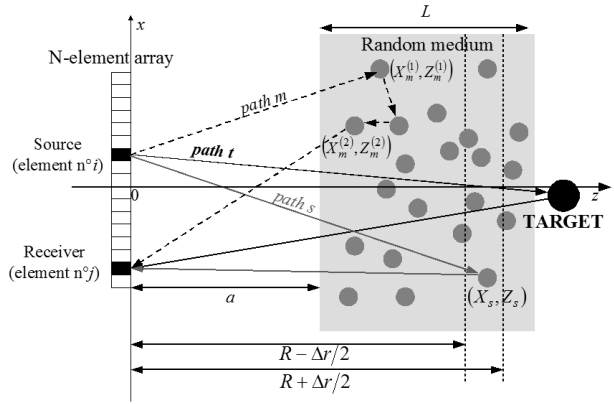}
\caption{\label{fig:setup}
Experimental setup. A 125-element array is placed in front of a random medium at
a distance $a$. The whole setup is immersed in a water tank. The inter-element response $k_{ij}(T,f)$, around the time of flight $T$ and at the frequency $f$, is measured. It contains contributions of single and multiple scattering paths whose lengths belong to the interval $[R-\Delta r/2;R+\Delta r/2]$, where $R=cT/2$ and $\Delta r = c\Delta t /2$. Examples of a single scattering path (labelled $s$, grey line) and of a multiple-scattering path (labelled $m$, dashed black line) is drawn. $(X_s,Z_s)$ are the coordinates of the scatterer involved in path $s$. $(X^{(1)}_m,Z^{(1)}_m)$ and $(X^{(2)}_m,Z^{(2)}_m)$ are the coordinates the first and last scatterers along path $m$. The path $t$ represents the single scattering path associated with the target (continuous black line).}
\end{figure}

\begin{figure}[htbp] 
\includegraphics{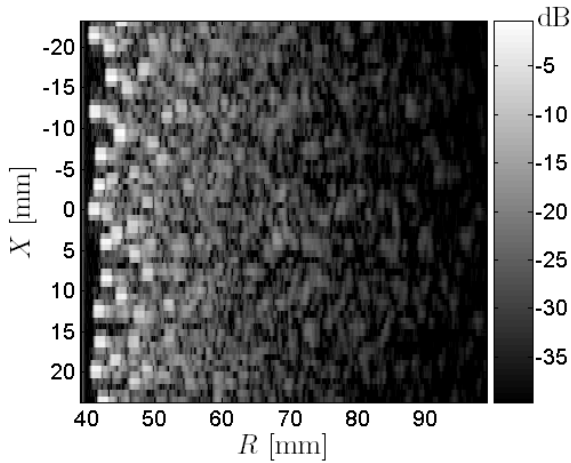}
\caption{\label{fig:echo_temp}Standard echographic image of the inspected medium obtained with focused and adaptative beamforming both at emission and reception. The image is normalized with its maximum and shown in dB.}
\end{figure}

The experiment takes place in a water tank. We use an N-element ultrasonic array ($N=125$) with a 3 MHz central frequency and a 2.5-3.5 MHz bandwidth; each array element is 0.39 mm in size and the array pitch $p$ is 0.417 mm. The sampling frequency is 20 MHz. The array is placed in front of the medium of investigation, at a distance $a=40$ mm. It consists of parrallel steel rods (longitudinal wave velocity $c_L=5.9$ mm/$\mu$s, transverse wave velocity $c_T=3$ mm/$\mu$s, radius $0.4$ mm, $\rho=7.85$ kg/L) randomly distributed with a concentration $n=12$ rods/cm$^2$. The frequency-averaged scattering mean-free path $l_e$ is $7.7\pm 0.3$ mm for this medium between 2.5 and 3.5 MHz \cite{derode4}. 
The slab thickness is $L=20$ mm. An air-filled steel cylinder with diameter 15 mm is placed behind the scattering slab. Our aim is to detect this echogene target. Note that the single scattered wave associated to the target (path $t$ in Fig.\ref{fig:setup}) has to travel more than five scattering mean free paths through the random medium. Its intensity is roughly divided by $\exp \left ( -2L/l_e \right) \sim 180$ as it traverses twice the scattering slab. Multiple scattering, in addition to aberration effects induced by the slab, make the detection of the target very difficult with classical imaging techniques. This is highlighted by the echographic image in Fig.\ref{fig:echo_temp}. The first rows of
scatterers in the slab are clearly visible. Beyond a depth of typically one mean free path ($\sim 5-10$ mm), the image displays a
speckle pattern without connection with the medium's reflectivity. The target, which should
be visible in Fig.\ref{fig:echo_temp} around $R= 70$ mm, is not detected by classical echography.

We now turn to the acquisition of the inter-element matrix (see Fig.~\ref{fig:setup}). A $2.5$-$\mu s$-long sinusoidal burst of frequency 3 MHz is emitted from transducer $i$ into the scattering sample. The backscattered wave is recorded with the $N$ transducers of the same array. The operation is repeated for the $N$ emitting transducers. The impulse response between transducers $i$ and $j$ is noted $h_{ij}(t)$. An $N \times N$ response matrix $\mathbf{H}(t)$ whose elements are the $N^2$ impulse responses $h_{ij}(t)$ is thus obtained. Because of reciprocity, $h_{ij}(t)=h_{ji}(t)$ and $\mathbf{H}(t)$ is symmetric. In the following, we take as the origin of time $t=0$, the instant when the source emits the incident wave.

A short-time Fourier analysis of the impulse response matrix $\mathbf{H}$ is achieved. The time signals $h_{ij}(t)$ are truncated into successive time windows : $k_{ij}(T,t)=h_{ij}(T-t)W_R(t)$ with $W_R(t)=1 \; \text{for} \; t\in[-\Delta t / 2 \; , \; \Delta t / 2]$, $W_R(t)=0$ elsewhere.
The value of $\Delta t$ is chosen so that signals associated with the same scattering event(s) within the medium arrive in the same time window \cite{aubry}. Actually, the choice of $\Delta t$ is particularly important for single scattering signals, if one wants to detect scatterers properly with the DORT method. In our experimental configuration, we obtain a value $\Delta t \simeq 11 \mu s$. For each value of time $T$, the $k_{ij}$ form a matrix $\mathbf{K}$. A Fourier analysis is achieved by means of a discrete Fourier transform (DFT) and gives a set of response matrices $\mathbf{K}(T,f)$ at time $T$ and frequency $f$. 

\section{\label{sec:coherence_ss}Single and multiple scattering contributions}
\begin{figure}[htbp] 
\includegraphics{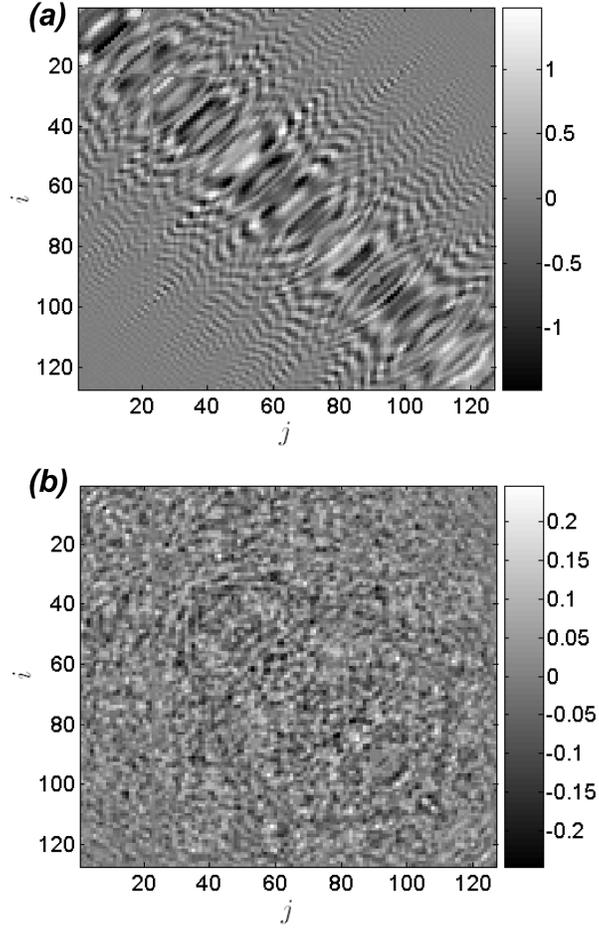}
\caption{\label{fig:ss_ms}Real part of matrix $\mathbf{K}$ at $f=3$ MHz and different arrival times $T$: (a) when the single scattering contribution is predominant ($T=58.5$ $\mu s$); (b) when only multiple scattering occurs ($T=122.5$ $\mu s$).}
\end{figure}

As an example, Fig.\ref{fig:ss_ms} shows the real part of $\mathbf{K}$ at the central frequency
$f = 3$ MHz. At early times (Fig.\ref{fig:ss_ms}(a)), single scattering dominates: multiple scattered echoes have not yet reached the array. Fig.\ref{fig:ss_ms}(b) represents $\mathbf{K}$ at an arrival time larger than $2(a + L)/c$: at such times only multiple scattering can exist. $\mathbf{K}$ clearly exhibits a different
behavior in the single and multiple scattering regimes. Whereas multiple scattering results
in a seemingly random matrix $\mathbf{K}$, single scattered waves exhibit a deterministic coherence
along the antidiagonals of $\mathbf{K}$. The reason for this, and its consequences on the statistical properties of the singular values, have been discussed in \cite{aubry09,aubry}. We briefly recall the argument in this section.

The signals $k_{ij}(T,f)$ can be written as the sum of a single scattering contribution $k_{ij}^S(T,f)$ and a multiple scattering contribution $k_{ij}^M(T,f)$
\begin{equation}
\label{eqn:ss_ms}
k_{ij}(T,f)=k_{ij}^S(T,f)+k_{ij}^M(T,f)
\end{equation}
Let us express both contributions.

The signals $k_{ij}^S(T, f)$ at a time $T$ and frequency $f$ correspond to the sum of partial waves that reach the array in the time window $[T-\Delta t/2 ;T+\Delta t/2]$. The ``isochronous volume'' is defined as the ensemble of points that contribute to the backscattered signal at a given time. It is formed by a superposition of ellipses whose foci are transmitter $i$ and receiver $j$. In a far-field configuration, we can approximate the isochronous volume by a slab of thickness $\Delta r=c\Delta t$, located at a distance $R=cT/2$ from the array and parallel to it (see Fig.\ref{fig:setup}). For simplicity but without loss of generality, we also assume that the reflectors as well as the array elements are point-like. In a 2D configuration, under the paraxial approximation, $k^S_{ij}(T,f)$ can be expressed as
\begin{equation}
\label{eqn:eq_ss_signal}
k_{ij}^S(T,f) \propto \frac{\exp \left ( j 2kR \right )}{R}\sum_{d=1}^{N_d} A_d \exp \left [ j k \frac{ \left (x_i -X_d \right )^2}{2R} \right ] \exp \left [ j k \frac{ \left (x_j -X_d \right )^2}{2R} \right ]
\end{equation}
where $k=2 \pi f / c$ is the wave number in the surrounding medium, ${x_i=(i-N/2)p}$ is the coordinate along the array. $X_d$ is the transverse position of the $d^{th}$ scatterer which contributes to the backscattered wave at time $T$, the amplitude $A_d$ depending on its reflectivity.  Both $A_d$ and $X_d$ are considered random. Note that $j=\sqrt{-1}$ in Eq.\ref{eqn:eq_ss_signal} and has not to be mixed up with the subscript $j$ which denotes the receiver index.

As to the multiple scattering contribution, $k_{ij}^M(T,f)$ also correspond to a sum of partial waves that reach the array in the time window $[T-\Delta t/2 ;T+\Delta t/2]$. They are associated with multiple scattering paths whose length belongs to the interval $[R-\Delta r/2;R+\Delta r/2]$, where $R=cT/2$ and $\Delta r=c\Delta t/2$. An example of such a path is drawn in Fig.\ref{fig:setup}. In a 2D configuration, under the paraxial approximation and assuming point-like transducers and scatterers, $k_{ij}^M(T,f)$ can be expressed as
\begin{equation}
\label{eqn:eq_ms_signal}
k^M_{ij}(T,f) \propto \sum_{q=1}^{N_q} B_q   \frac{\exp \left [ j k \left ( Z^{(1)}_q + Z^{(2)}_q \right ) \right ]}{\sqrt{Z^{(1)}_qZ^{(2)}_q}} \exp \left[ j k \frac{ \left (x_i -X_q^{(1)} \right )^2}{2Z^{(1)}_q} \right ] \exp \left [ j k \frac{ \left (x_j -X_q^{(2)} \right )^2}{2Z^{(2)}_q} \right ]
\end{equation}
The index $q$ denotes the $q^{th}$ path which contributes to the signal received at time $T$. $\left (X_q^{(1)},Z_q^{(1)}\right )$ and $\left (X_q^{(2)},Z_q^{(2)}\right )$ are respectively the coordinates of the first and last scatterers along the path $q$. $B_q$ is the complex amplitude associated with path $q$, from the first scattering event at $\left (X_q^{(1)},Z_q^{(1)}\right )$ until the last one at $\left (X_q^{(2)},Z_q^{(2)}\right )$.

At this stage, the theoretical expression of $k_{ij}^S(T,f)$ given in Eq.\ref{eqn:eq_ss_signal} does not display any obvious coherence: $k_{ij}^S(T,f)$ corresponds to a sum of partial waves which are independent of each other since the distribution of scatterers is assumed random. One can try to express $k_{ij}^S(T,f)$ as a function of $(x_i-x_j)$ and $(x_i+x_j)$ which corresponds to a change of coordinates in Eq.\ref{eqn:eq_ss_signal}:
\begin{equation}
\label{eqn:eq_ss_signal_2}
k_{ij}^S(T,f) \propto  \underbrace { \frac{\exp \left ( j 2kR \right )}{R} \exp \left [ j k \frac{ \left (x_i -x_j \right )^2}{4R} \right ] }_{\mbox{deterministic term}}  \underbrace{ \sum_{d=1}^{N_d} A_d  \exp \left [ j k \frac{ \left (x_i + x_j - 2X_d \right )^2}{4R} \right ] }_ {\mbox{random term}} 
\end{equation}
The term before the sum in Eq.\ref{eqn:eq_ss_signal_2} does not depend on the distribution of scatterers, it is deterministic; on the contrary, the term on the right is random. This special feature of single scattering signals manifests itself as a particular coherence along the antidiagonals of the matrix $\mathbf{K}$, as depicted by Fig.\ref{fig:ss_ms}(a). Indeed, along each antidiagonal, \textit{i.e} for couples of transmitter($i$) and receiver($j$) such as $i + j$ is constant, the random term of Eq.\ref{eqn:eq_ss_signal_2} is also constant,  for any given realization of disorder.
Thus, there is a deterministic phase relation between coefficients of $\mathbf{K}$ located on the same antidiagonal. It can be expressed in the following way :
\begin{equation}
\label{eqn:eq_ss_signal_3}
\beta_{m}=\frac{k^S_{i-m,i+m}(T,f)}{k^S_{ii}(T,f)} =  \exp \left [ j k \frac{ \left (m p \right )^2}{R} \right ]
\end{equation}
This is no longer true in the multiple scattering regime, since $k^M_{ij}$ cannot be factorized so simply.
Note that the parabolic phase dependence along each antidiagonal of $\mathbf{K^S}$ should be weighted
by an attenuation term, decreasing with $\left | x_i-x_j \right |$, in order to incorporate the directivity of
transducers. Thus Eq.\ref{eqn:eq_ss_signal_2} is not rigorously true; yet for simplicity, we will neglect this attenuation term in the following.

\section{\label{sec:anti_filtering}Single scattering filter (SSF)}

Now that we have explained the deterministic coherence of single scattering signals along the antidiagonals of the array response matrix $\mathbf{K}$, we can take advantage of this special feature to extract the single scattering contribution from the multiple scattering background. Once the set of matrices $\mathbf{K}(T,f)$ are measured, the separation between single and multiple scattering contributions is achieved according to the following steps:
\begin{itemize}
\item Rotation of each matrix $\mathbf{K}$ and construction of two sub-matrices $\mathbf{A_1}$ and $\mathbf{A_2}$.
\item Filtering of matrices $\mathbf{A_1}$ and $\mathbf{A_2}$. Two new matrices $\mathbf{A_1^F}$ and $\mathbf{A_2^F}$ are obtained.
\item Construction, from $\mathbf{A_1^F}$ and $\mathbf{A_2^F}$, of the filtered matrices $\mathbf{K^F}$ containing the single scattering signals.
\end{itemize}
In the following subsections, we explain in details the matrix operations performed at each step.

\subsection{\label{subsec:rot_anti}Matrix rotation}
\begin{figure}[htbp] 
\includegraphics{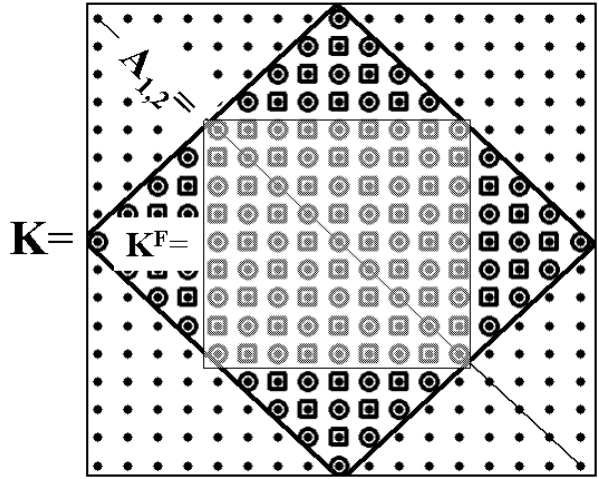}
\caption{\label{fig:rotation}Principle of the data rotation by taking the example of a matrix $\mathbf{K}$ of dimension $N=17$. The black points denote the whole elements $k_{ij}$ of $\mathbf{K}$. The columns of matrices $\mathbf{A_1}$ and $\mathbf{A_2}$ contain the antidiagonals of $\mathbf{K}$. Circles and squares represent respectively elements of $\mathbf{A_1}$ and $\mathbf{A_2}$. After filtering $\mathbf{A_1}$ and $\mathbf{A_2}$, the filtered matrix $\mathbf{K^F}$ is obtained and corresponds to elements contained in the central square. Its dimension is $(2M-1)\times (2M-1)$ ($M=5$ here).}
\end{figure}

A rotation of matrix elements is achieved as depicted in Fig.\ref{fig:rotation}. It consists in building two matrices $\mathbf{A_1}$ and $\mathbf{A_2}$ from the matrix $\mathbf{K}$:
\begin{eqnarray}
\label{eqn:construction_A1}
\mathbf{A_1}   =  \left [ a_{1uv} \right ]  \, & \mbox{of dimension }&(2M-1) \times (2M-1) \mbox{,}\nonumber \\
& \mbox{such that } & a_1[u, v]  = k[u + v -1, v - u + 2M -1] \\
\mathbf{A_2} =  \left [ a_{2uv} \right ]  \, & \mbox{of dimension }&(2M-2) \times (2M-2)\mbox{,}\nonumber \\
& \mbox{such that } & a_2[u, v]  = k[u + v , v - u + 2M -1]
\label{eqn:construction_A2}
\end{eqnarray}
with $M=(N+3)/4$. Here $N=125$ and so $M=32$ is an even number.
The matrices $\mathbf{A_1}$ and $\mathbf{A_2}$ contain the whole antidiagonals of $\mathbf{K}$ (see Fig.\ref{fig:rotation}). Therefore the coherence of single scattering signals now manifests itself along the columns of $\mathbf{A_1}$ and $\mathbf{A_2}$. In the next subsection, we will no longer make the difference between matrices $\mathbf{A_1}$ and $\mathbf{A_2}$ because they are filtered in the same way. They will be called indifferently $\mathbf{A}$. $L$ is the dimension of $\mathbf{A}$. For matrix $\mathbf{A_1}$, $L=2M -1$; for matrix $\mathbf{A_2}$, $L=2M -2$. Because of spatial reciprocity, $\mathbf{K}$ is symmetric ($k_{ij}=k_{ji}$). Thus, $\mathbf{A}$ also exhibits a symmetry: each line of its upper part is identical to a line of its lower part. The symmetry axis is shown as a black line in Fig.\ref{fig:rotation} and corresponds to the diagonal of the matrix $\mathbf{K}$. So, each column of the matrix $\mathbf{A}$ contains only $M$ independent coefficients, even if its dimension $L$ is superior to $M$. This fact will be crucial when the gain in signal-to-noise ratio will be assessed.

\subsection{\label{subsec:filter_anti}Filtering of matrix A}

The matrix $\mathbf{A}$ is the sum of two matrices $\mathbf{A^S}$ and $\mathbf{A^M}$, which correspond respectively to the single and multiple scattering contributions
\begin{equation}
\label{eqn:ss_ms_anti}
\mathbf{A}=\mathbf{A^S}+\mathbf{A^M}
\end{equation}
The rotation of data can be described as the following change of coordinates $(x_i,x_j)\rightarrow(y_u,y_v)$:
\begin{equation}
\label{eqn:sub_var}
y_u =\frac{x_i - x_j}{\sqrt{2}}\; \mbox {  and  } \; y_v =\frac{x_i + x_j}{\sqrt{2}}
\end{equation}
In this new basis, Eq.\ref{eqn:eq_ss_signal_2} becomes
\begin{equation}
\label{eqn:eq_ss_signal_anti}
a_{uv}^S(T,f) \propto  \underbrace { \frac{\exp \left ( j 2kR \right )}{R} \exp \left [ j k \frac{  y_u^2}{2R} \right ] }_{\mbox{deterministic term}} \times  \underbrace{\Gamma_v }_ {\mbox{random term}} 
\end{equation}
where $ \Gamma_v = \sum_{d=1}^{N_d} A_d  \exp \left [ j k \frac{ \left (y_v - \sqrt{2} X_d \right )^2}{2R} \right ]$. Each column of the matrix $\mathbf{A^S}$ exhibits a known dependence as a function of index $u$. On the contrary, the multiple scattering contribution (Eq.\ref{eqn:eq_ms_signal}) cannot be factorized in this way. Even after rotation, the random feature remains along the columns and the lines of matrix $\mathbf{A^M}$.

The extraction of single scattering signals can be achieved by projecting the columns of the matrix $\mathbf{A}$ on the ``characteristic space'' of single scattering, generated by the vector $\mathbf{S}$ whose coordinates are
\begin{equation}
\label{eqn:vector_s}
s_u = \exp \left [ j k \frac{  y_u^2}{2R} \right ]L^{-1/2} 
\end{equation}
The factor $L^{-1/2} $ ensures the normalization of $\mathbf{S}$. The result $\mathbf{P}$ of this projection is
\begin{equation}
\label{eqn:vector_P}
\mathbf{P}=\mathbf{S}^{\dag}\mathbf{A}
\end{equation}
whose coordinates are
\begin{eqnarray}
\label{eqn:coord_P}
p_v&=&\sum_{u=1}^L s_u^*a_{uv}= \sum_{u=1}^L s_u^*a^S_{uv} + \sum_{u=1}^L s_u^*a^M_{uv}\\
&=& \sqrt{L}\frac{\exp \left ( j 2kR \right )}{R}\Gamma_v + \sum_{u=1}^L s_u^*a^M_{uv}
\end{eqnarray}
The residual term $\sum_{u=1}^L s_u^*a^M_{uv}$ corresponds to the projection of multiple scattering signals on the vector $\mathbf{S}$. The filtered matrix $\mathbf{A^F}$ is obtained by multiplying the column vector $\mathbf{S}$ by the line vector $\mathbf{P}$
\begin{equation}
\label{eqn:Af}
\mathbf{A^F}=\mathbf{S}\mathbf{P}=\mathbf{S}\mathbf{S}^{\dag}\mathbf{A}
\end{equation}
The elements of $\mathbf{A^F}$ are:
\begin{equation}
\label{eqn:Af_coord}
a^F_{uv}= \frac{\exp \left ( j 2kR \right )}{R} \exp \left [ j k \frac{  y_u^2}{2R} \right ] \Gamma_v + s_u \sum_{u'=1}^L s_{u'}^*a^M_{u'v}
\end{equation}
The first term on the right-hand side of Eq.\ref{eqn:Af_coord} is strictly equal to the single scattering component (Eq.\ref{eqn:eq_ss_signal_anti}). Finally, we obtain
\begin{equation}
\label{eqn:Af_coord_2}
a^F_{uv}= a^S_{uv} + s_u \sum_{u'=1}^L s_{u'}^*a^M_{u'v}
\end{equation}
Eq.\ref{eqn:Af_coord_2} can be written under a matrix formalism:
\begin{equation}
\label{eqn:Af_2}
\mathbf{A^F}=\underbrace{\mathbf{A^S}}_{\mbox{Single scattering}}+ \underbrace{\mathbf{S}\mathbf{S}^{\dag}\mathbf{A^M}}_{\mbox{Residual noise}}
\end{equation}
The matrix $\mathbf{A^F}$ contains the single scattering contribution ($\mathbf{A^S}$) as wanted. But it also contains a residual term due to multiple scattering ($\mathbf{S}\mathbf{S}^{\dag}\mathbf{A^M}$). This term persists because multiple scattering signals are not stricly orthogonal to the characteristic space of single scattering, generated by the vector $\mathbf{S}$. The filtering of the single scattering contribution is not perfect. Nevertheless, the typical amplitude of the residual noise can be assessed. Since each column of $\mathbf{A}$ contains $M$ independent coefficients, the filtering process decreases the multiple scattering contribution by a factor $\sqrt{M}$. The single scattering contribution remaining unchanged, the gain in signal-to-noise ratio (in amplitude), or rather the gain in ``single-scattering-to-multiple-scattering'' ratio, is of $\sqrt{M}$. 

\subsection{\label{subsec:buiding_Kf}The filtered matrix $\mathbf{K^F}$}
Once the matrices $\mathbf{A}_1$ and $\mathbf{A}_2$ are filtered, an inter-element filtered matrix $\mathbf{K^F}$, of dimension $(2M-1)\times(2M-1)$, is built (see Fig.\ref{fig:rotation}) with a change of coordinates, back to the original system:
\begin{itemize}
\item if $(i-j)/2$ is an integer,\\ then, $k^F[i, j]= a_1^F [(i - j)/ 2 + M,(i + j)/ 2]$
\item if $(i-j)/2$ is not an integer,\\ then, $ k^F[i, j]= a_2^F [(i - j -1)/ 2 + M, (i + j -1)/ 2]$
\end{itemize}
In the following, $\mathbf{K^0}$ will denote the matrix that would have been obtained if no filtering had been performed. $\mathbf{K^0}$ has the same dimensions as $\mathbf{K^F}$, and simply contains the central elements of $\mathbf{K}$.

\subsection{\label{subsec:truncation}Illustration of the single scattering filter (SSF)}

\begin{figure}[htbp] 
\includegraphics[width=85mm]{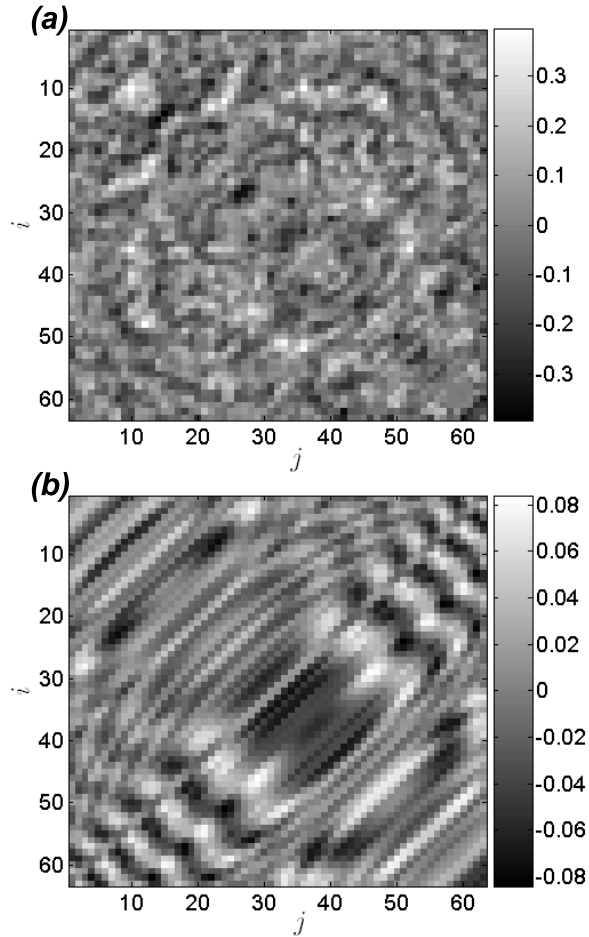}
\caption{\label{fig:fig5}Results given by the SSF at time $T=94.5$ $\mu$s and frequency $f=2.7$ MHz. (a) Real part of matrix $\mathbf{K^0}$. (b) Real part of filtered matrix $\mathbf{K^F}$.}
\end{figure}

As an example, Fig.\ref{fig:fig5} illustrates the action of the SSF on experimental data. Matrices $\mathbf{K^0}$(Fig.\ref{fig:fig5}(a)) and $\mathbf{K^F}$(Fig.\ref{fig:fig5}(b)) are shown at frequency $f=2.7$ MHz and time $T=94.5$ $\mu s$. This arrival time is the one expected for the target echo. Whereas the matrix $\mathbf{K^0}$ seems random, the filtered matrix $\mathbf{K^F}$ displays a deterministic coherence along its antidiagonals. 
From these data, an additional operation is needed to build the image of the medium or rather, to detect and image the target placed behind the scattering slab. To that aim, several imaging techniques are compared in the next section.  

\section{\label{detect_imag}Detection and imaging of the target}

In this section, we apply the filtering procedure described above to the detection
and imaging of a target embedded in a scattering medium. At a given frequency, two imaging techniques are compared: focused beamforming (equivalent to echography in the frequency domain) and the DORT method. As we will see, there is no interest in combining the SSF with focused beamforming (FB). But its combination with the DORT method provides excellent results.

\subsection{\label{subsec:frequential_ultrasono}Focused beamforming (FB)}

\begin{figure}[htbp] 
\includegraphics{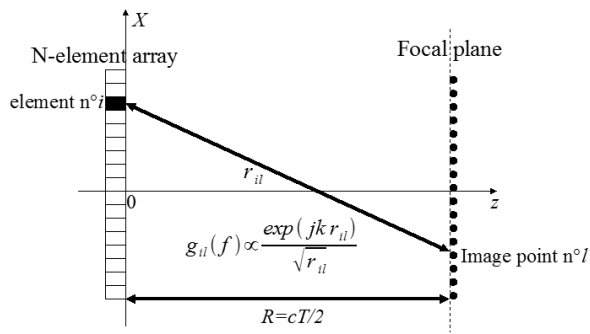}
\caption{\label{fig:fig6}Principle of FB. The focal plane is parallel to the array and located at depth $R=cT/2$ from it. It is discretized with a sampling period equal to the array pitch. The distances $r_{il}$ are much larger than the wavelength.}
\end{figure}

\begin{figure}[htbp] 
\includegraphics{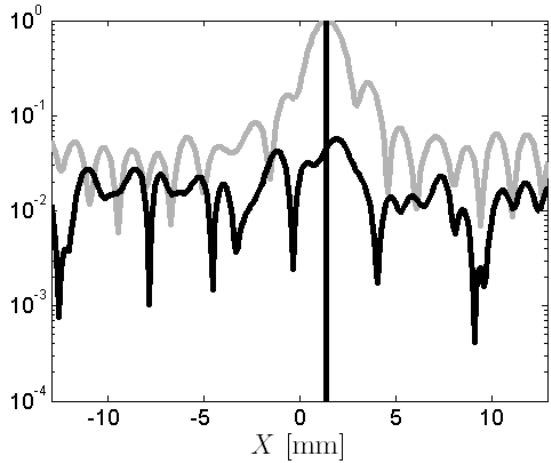}
\caption{\label{fig:fig7} Image obtained by FB at time $T=94.5$ $\mu$s and frequency $f=2.7$ MHz (black line). The ideal image obtained without the scattering slab is in grey. The vertical black line indicates the position of the target.}
\end{figure}

The simplest way to image the target is to achieve a direct backpropagation of the measured signals $\mathbf{K}(T,f)$, for a given time and frequency couple. The focal plane is parallel to the array and located at depth $R = cT/2$; it is discretized in a set of points. The backpropagation algorithm is based on the Born approximation. It consists first in calculating the propagation operator $\mathbf{G}$, whose elements are the spatial Green functions $g_{il}$ between the $i^{\mbox{\small th}}$ array element and the $l^{\mbox{\small th}}$ point in the focal plane, as shown in Fig.\ref{fig:fig6}. The medium is considered as homogeneous with a wave celerity $c$ equal to that of the surrounding medium. 
At a given time $T$ (corresponding to depth $R = cT/2$) and frequency $f$, the final image is a vector $\mathbf{I}$, the absolute value of the backpropagated wave field, which can be plotted as a function of $X$, the transverse coordinate in the focal plane:
\begin{equation}
\label{eqn:echo_intensity}
\mathbf{I}=\left | \mathbf{G}^{\dag}\mathbf{K^0G}^* \right |
\end{equation}
This backpropagation algorithm is the equivalent of echography in the frequency domain, with a poorer temporal resolution due to the duration $\Delta t$ of the time-windows. In the following, we will refer to this imaging technique as ``focused beamforming''(FB). FB has the same drawbacks as classical echography (Fig.\ref{fig:echo_temp}), particularly the presence of speckle which hides the target. We can point out its inability to detect the target by considering the image obtained at time $T=94.5$ $\mu$s and $f=2.7$ MHz (see Fig.\ref{fig:fig7}). This arrival time corresponds to the target depth, and 2.7 MHz is the frequency for which the mean-free path of the slab is the largest (\textit{i.e.} multiple scattering is the weakest) \cite{derode4}. The presence of the scattering slab seriously degrades the image. Two peaks seem to arise but neither of them is located at the expected position. 

The resulting image $\mathbf{I}$ (Eq.\ref{eqn:echo_intensity}) could also be averaged over the whole frequency domain (the result would be comparable to Fig.\ref{fig:echo_temp}), or on specific frequency bands for which the detection is more likely. Indeed, it is possible to establish a detection criterion based on speckle statistics, for a given probability of false alarm. This will be done, as well as for other techniques, in Sec.\ref{sec:detect_crit}.

In order to improve the results provided by FB, one could think of applying Eq.\ref{eqn:echo_intensity} to the filtered matrix $\mathbf{K^F}$ instead of the raw matrix $\mathbf{K^0}$. Yet it can be shown (Appendix \ref{app:non_complement}) that this  would not change the result. A short interpretation can be given. FB relies on the fact that single scattering signals will add up coherently as long as they come from a focal point at depth $R = cT/2$ (axial focusing) and the desired transverse position $X$ (lateral focusing). The SSF also enhances single scattering signals associated with scatterers located around $R = cT/2$, but independently from their transverse position $X$. Now, if we build the echographic image (Eq.\ref{eqn:echo_intensity}) from the filtered signals, there is a redundancy in the choice of depth $R$; consequently, the SSF does not bring anything when it is followed by FB. The two techniques are not complementary. A more rigorous demonstration is given in Appendix \ref{app:non_complement}. 

\subsection{DORT applied to $\mathbf{K^0}$}

DORT \cite{prada,prada2} consists in achieving the singular value decomposition (SVD) of the array response matrix before imaging the medium: 
\begin{equation}
\label{eqn:dort}
\mathbf{K^0} = \mathbf{U^0} \mathbf{\Lambda^0} \mathbf{V^{0\dag}}
\end{equation}
where $\mathbf{\Lambda^0}$ is a diagonal matrix containing the real positive singular values $\lambda^0_i$ in a decreasing order ($\lambda^0_1>\lambda^0_2>...> \lambda^0_M$). $\mathbf{U^0}$ and $\mathbf{V^0}$  are unitary matrices containing the normalized singular vectors $\mathbf{U^0_i}$ and $\mathbf{V^0_i}$. Under the single scattering approximation, each scatterer of the medium is mainly linked with one singular space associated to a non-zero singular value $\lambda_i^0$. The corresponding singular vector $\mathbf{V_i^0}$ is the signal to apply to the array in order to focus on the corresponding scatterer. Thus, the numerical backpropagation of the singular vectors allows to image each detected scatterer. The image provided by DORT is a vector $\mathbf{I^0_i}(T,f)$ which corresponds to the absolute value of the backpropagated wave field
\begin{equation}
\label{eqn:dort_image}
\mathbf{I^0_i}=\lambda_i^0 \left | \mathbf{V_i^0} \mathbf{G}^* \right |
\end{equation}
$\mathbf{I^0_i}(T,f)$ represents the backpropagated image at time $T$ and frequency $f$ of the $i^{th}$ singular vector $\mathbf{V_i^0}(T,f)$.

\begin{figure}[htbp] 
\includegraphics{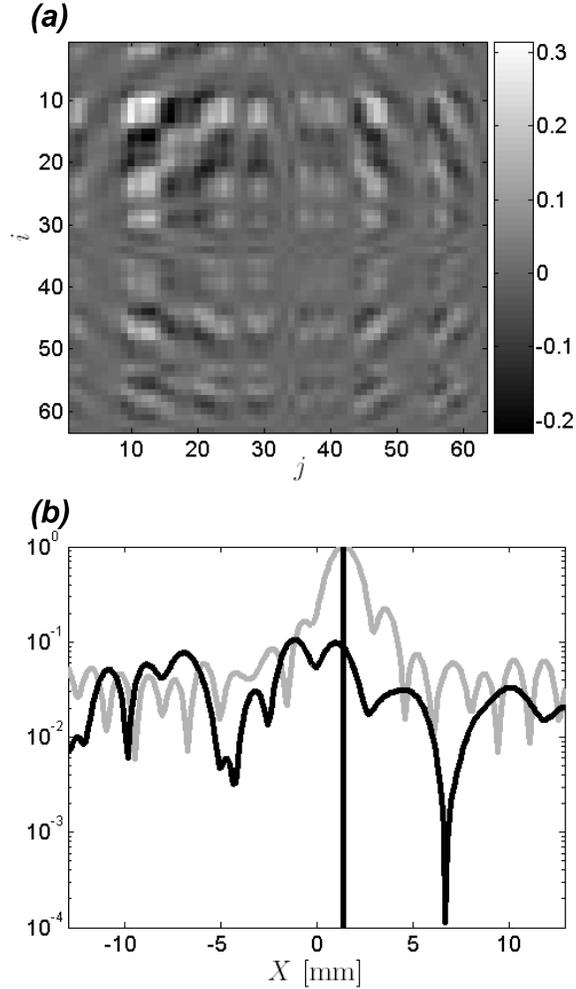}
\caption{\label{fig:fig8} (a) Real part of the first singular space $\lambda^0_1 \mathbf{U_1^0}\mathbf{V_1^{0\dag}}$ of $\mathbf{K^0}$ obtained at time $T=94.5$ $\mu$s and frequency $f=2.7$ MHz. (b) Image obtained by backpropagation of the first singular vector $\mathbf{V_1^0}$ at the same time-frequency couple. The DORT image (black line) is compared with the ideal image obtained without the forest of rods (grey line). The vertical black line indicates the position of the target.}
\end{figure}

We would like the first singular space (linked to $\lambda^0_1$) to be associated with the target echo. However, the forest of rods in front of the target results in multiple scattering which hides the target echo. Its influence is illustrated in Fig.\ref{fig:fig8} which displays the result obtained with the DORT method at the expected arrival time for the target echo (94.5 $\mu$s), and at the frequency for which multiple scattering is at its weakest (2.7 MHz).
Fig.\ref{fig:fig8}(a) represents the real part of the first singular space $\lambda^0_1 \mathbf{U^0_1} \mathbf{V^{0\dag}_1}$ of the matrix $\mathbf{K^0}$ (see Fig.\ref{fig:fig5}(a)). It does not display the feature of a single scattered echo (\textit{i.e} concentric circles centered around the target position like in Fig.\ref{fig:fig9}(a)). Multiple scattering results in a random matrix $\mathbf{K^0}$ (see Fig.\ref{fig:fig5}(a)) whose singular spaces are random, without connection with the direct echoes of scatterers located in the isochronous volume. The corresponding image $\mathbf{I^0_1}$ obtained by backpropagation of the singular vector $\mathbf{V^0_1}$ is shown in Fig.\ref{fig:fig8}(b). No peak is observed around the target location.
\subsection{DORT applied to the filtered matrix $\mathbf{K^F}$}

\begin{figure}[htbp] 
\includegraphics{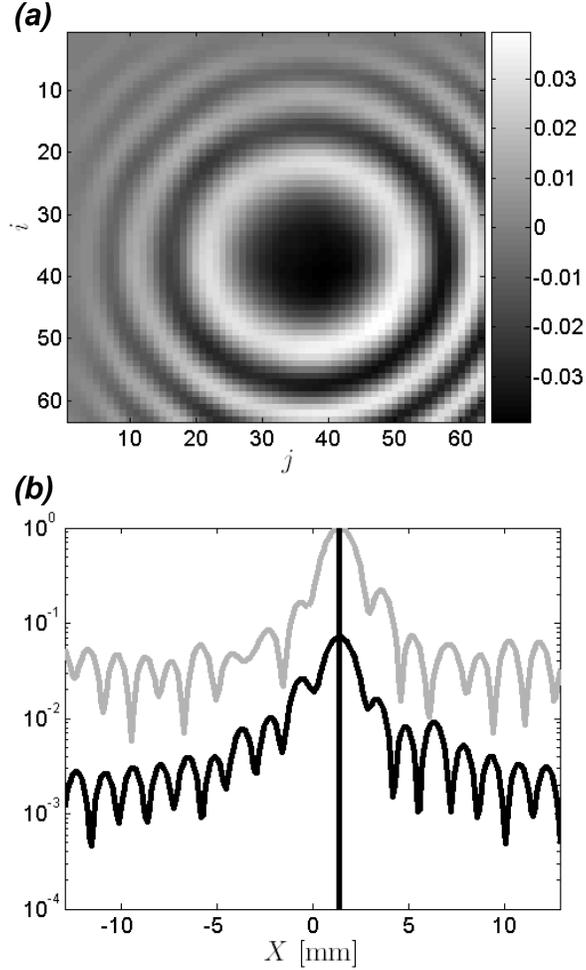}
\caption{\label{fig:fig9} (a) Real part of the first singular space $\lambda_1^F \mathbf{U_1^F}\mathbf{V_1^{F\dag}}$ of $\mathbf{K^F}$ obtained at time $T=94,5$ $\mu$s and frequency $f=2,7$ MHz. (b) Image obtained by backpropagation of the first singular vector $\mathbf{V_1^F}$ at the same time-frequency couple. The image (black line) is compared with the ideal image obtained without the forest of rods (grey line). The vertical black line indicates the position of the target.}
\end{figure}

Here we combine the DORT method with the SSF described in Sec.\ref{sec:anti_filtering} (SSF+DORT approach). The procedure is the same as the one described in the previous subsection, except that $\mathbf{K^0}$ is replaced by $\mathbf{K^F}$. Fig.\ref{fig:fig9} illustrates the success of this combination. 
The filtered matrix $\mathbf{K^F}$ (see Fig.\ref{fig:fig5}(b)) already shows a possible feature of a single scattered echo coming from the target. Nevertheless, it is still perturbed by a residual multiple scattering contribution. Once the SVD of  $\mathbf{K^F}$ is achieved ($\mathbf{K^F}=\mathbf{U^F} \mathbf{\Lambda^F} \mathbf{V^{F\dag}}$), the first singular space $\lambda^F_1 \mathbf{U^F_1} \mathbf{V^{F\dag}_1}$, shown in Fig.\ref{fig:fig9}(a), clearly exhibits the feature of the single scattered echo coming from the target. The backpropagation of the singular vector $\mathbf{V^{F}_1}$ is shown in Fig.\ref{fig:fig9}(b). The image clearly displays a peak at the target position, with a spatial resolution that is comparable to the free-space situation! Yet the peak amplitude is lower, since the intensity of the coherent wave coming from the target has undergone an attenuation of $\sim \exp(2L/l_e)$ due to the forest of rods. 

The comparison of Figs.\ref{fig:fig7}, \ref{fig:fig8} \& \ref{fig:fig9} illustrates the success of our approach. The SSF eliminates a major part of the multiple scattering contribution. It allows the DORT method to extract properly the target echo, which was not possible with classical imaging techniques. However, these results are obtained for a time-frequency couple chosen arbitrarily. An imaging procedure must work blindly, without knowing the depth of the target or the frequency band in which the medium is less scattering. The detection of the target has to be systematized in order to select automatically the arrival time(s) and frequency band(s) for which the target is detected and can be imaged. Moreover, multiple scattering signals can generate false alarms because of speckle fluctuations that one can wrongly attribute to the presence of a strong reflector in the medium. Thus, whatever the imaging procedure, a rigorous detection criterion has to be established in order to discriminate artifacts and compare the different techniques on a common basis. This is done in the next section.

\section{\label{sec:detect_crit}Detection criteria}

At a given time $T$ and frequency $f$, a target will be detected if the observed quantity is above a certain threshold. In the case of the DORT method, the detection criterion will be applied to the first singular value $\lambda_1$. In the case of FB, the relevant variable is the maximum of the image $\mathbf{I}$ (Eq.\ref{eqn:echo_intensity}). Since the scattering medium is considered as one realisation of a random process, setting the detection criteria requires a statistical model for the probability density function of $\lambda_1$ and of $\mathbf{I}$. Then a probability of false alarm (PFA) is fixed, and the corresponding detection thresholds can be established for both methods, which allows to compare their results for the same PFA. 

The statistical behavior of the singular values $\lambda_i$ and of the echographic image $\mathbf{I}$ in the multiple scattering regime has to be known. To that aim, we have performed the same kind of experiments as described in Fig.\ref{fig:setup}, except that the target has been removed. The experimental procedure remains unchanged and a set of matrices $\mathbf{K^0}(T,f)$ and $\mathbf{K^F}(T,f)$ is obtained.

\begin{figure}[htbp] 
\center
\includegraphics{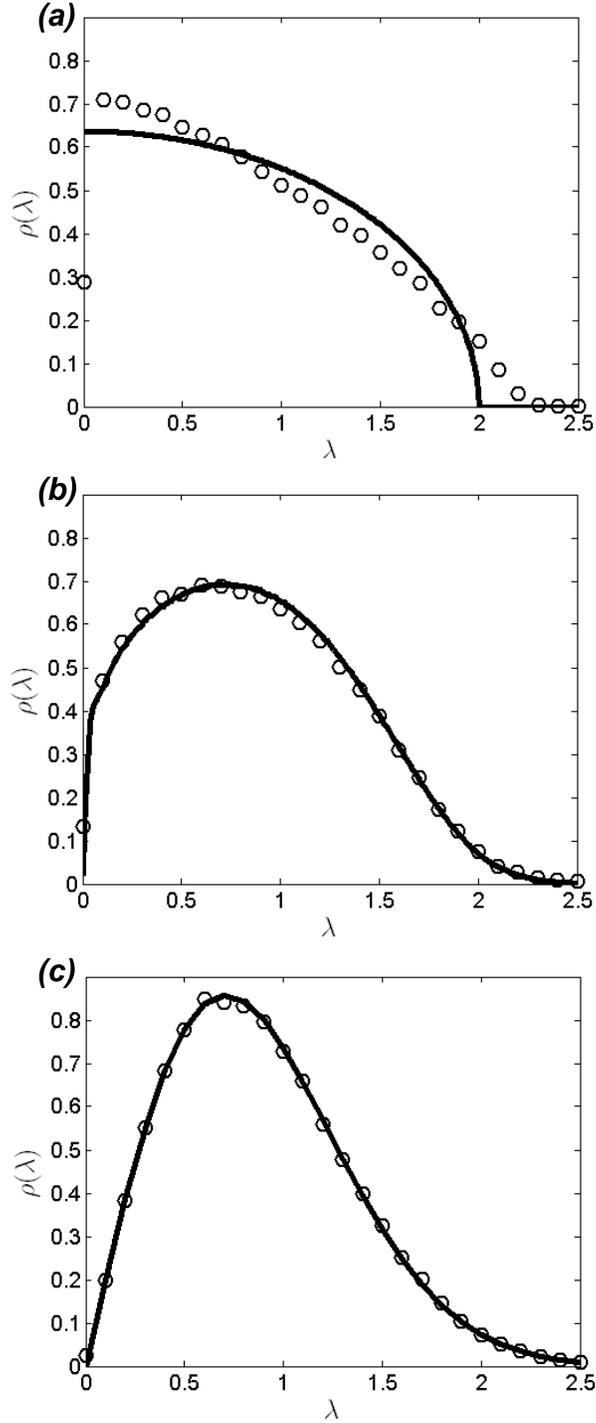}
\caption{\label{fig:fig10} (a) $\hat{\rho}_0(\lambda)$ (white disks) is compared with the quarter-circle law $\rho_{QC}(\lambda)$ (black line, Eq.\ref{eqn:eq5}). (b) $\hat{\rho}_F(\lambda)$ (white disks) is compared with the Hankel law $\rho_H(\lambda)$ (black line). (c) $\hat{\rho}_I(\lambda)$ (white disks) is compared with the Rayleigh law $\rho_R(\lambda)$ (black line, Eq.\ref{eqn:rayleigh_law_detection}).}
\end{figure}

We first consider the DORT method, and the statistical properties of the singular values of $\mathbf{K^0}$ and $\mathbf{K^F}$ in connection with random matrix theory (RMT), as discussed in recent papers \cite{aubry09,aubry}. Experimentally, before achieving the SVD, the matrices $\mathbf{K^0}$ and $\mathbf{K^F}$ of size $(2M-1) \times (2M-1)$ are truncated into matrices of size $M \times M$ by keeping only one element in two. This operation is needed in order to remove short-range correlations that may exist between adjacent entries \cite{aubry}. These correlations have an important influence on the distribution of singular values and removing them simplifies the problem. For the sake of simplicity, we will continue to note the truncated matrices $\mathbf{K^0}$ and $\mathbf{K^F}$. The SVD of these matrices is achieved and a set of $M$ singular values $\lambda_i^0(T,f)$ and $\lambda_i^F(T,f)$ is obtained at each time-frequency couple $(T,f)$. The singular values are normalized by their quadratic mean :
\begin{equation}
\label{eqn:eq2}
\tilde{\lambda}_i=\frac{\lambda_i}{\sqrt{\frac{1}{M}\sum_{p=1}^M \lambda_p^2}}
\end{equation}
This normalization allows to meet the hypothesis usually made in RMT which consists in assuming a variance of $\frac{1}{M}$ for matrix coefficients \cite{aubry}. Once this normalization is performed at each time-frequency couple, a histogram of dimensionless singular values is obtained by averaging over time $T$, frequency $f$ and rank $i$. Two estimators, $\hat{\rho}_0(\lambda)$ and $\hat{\rho}_F(\lambda)$, of the singular values distribution are finally obtained, respectively for matrices $\mathbf{K^0}$ and $\mathbf{K^F}$. The results are displayed in Fig.\ref{fig:fig10}.

In the multiple scattering regime, once short-range correlations are removed, we expect the matrix $\mathbf{K^0}$ to be random. In that case, RMT predicts that the distribution of singular values should follow the ``quarter-circle law'' (for $M>>1$) \cite{marcenko,tulino}
\begin{equation}
\label{eqn:eq5}
\rho_{QC}(\lambda)=
\left \{
\begin{array}{cl}
\frac{1}{\pi}\sqrt{4-\lambda^2} & \mbox{for}\,\, 0<\lambda<2 \\
0 & \mbox{elsewhere} 
\end{array}
\right.
\end{equation}
As pointed out by Fig.\ref{fig:fig10}(a), the experimental distribution of singular values $\hat{\rho}_0(\lambda)$ deviates from the quarter circle law $\rho_{QC}(\lambda)$. The reasons for that have been discussed in \cite{aubry}. When a detection threshold is fixed for the first singular value $\tilde{\lambda}_1^0$, we will use experimental data $\hat{\rho}_0$ rather than the theoretical quarter-circle law ($\rho_{QC}$).

Contrary to $\mathbf{K^0}$, the filtered matrix $\mathbf{K^F}$ is characterized by a deterministic phase relation along its antidiagonals. This kind of matrix has already been studied \cite{aubry}. $\mathbf{K^F}$ displays the same statistical properties as a Hankel random matrix. A Hankel matrix is a square matrix whose elements belonging to the same antidiagonal ($i+j=$ constant) are equal. In the literature, Bryc \textit{et al.} \cite{bryc} have proved, for normalized random Hankel matrices, the almost sure weak convergence of the distribution of singular values to a universal distribution of unbounded support $\rho_H(\lambda)$. In the following, the distribution $\rho_H(\lambda)$ will be referred to as the ``Hankel law''. To our knowledge, no analytical expression of the Hankel law has ever been found and only a numerical simulation can provide an estimate of $\rho_H(\lambda)$. In Fig.\ref{fig:fig10}(b), the experimental distribution of singular values of $\mathbf{K^F}$, $\hat{\rho}_F(\lambda)$, is compared to the Hankel law. The agreement between both curves is excellent. Thus, we will rely on the statistical behavior of Hankel random matrix, when a detection criterion is set on the first singular value $\tilde{\lambda}_1^F$.

We now consider the echographic image $\mathbf{I}(T,f)$ and build an estimator for its probability density function. Experimentally, the points where the reflectivity of the medium is estimated have to be chosen carefully. The image vector $\mathbf{I}(T,f)$ has to display independent coordinates. So, each one has to be associated with a different resolution cell. In pratice, we have considered the points located at the same transverse position as the transducers implied in the truncated matrices $\mathbf{K^0}$ and $\mathbf{K^F}$. The points of the image are separated by a distance $2p\simeq0.84$ mm larger than the size of the resolution cell $\frac{\lambda a}{D}\simeq0.76$ mm (with $D=(2M-1)p$ the array size). Each image $\mathbf{I}(T,f)$ contains M independent coordinates $I_l(T,f)$. These coordinates are normalized by their quadratic mean at each time $T$ and frequency $f$:
\begin{equation}
\label{eqn:renorm_coord_detection}
\tilde{I}_l(T,f)=\frac{I_l(T,f)}{\sqrt{\frac{1}{M}\sum_{p=1}^{M}I_p^2(T,f)}}
\end{equation}
Once this renormalization is performed, a histogram of the dimensionless image can be built, averaging over all time-frequency couples. The estimator $\hat{\rho}_I(\lambda)$ of the image probability density function is plotted in Fig.\ref{fig:fig10}(c).

In the multiple scattering regime, we expect $I_l$ to be the modulus of a gaussian complex random variable with zero mean and variance unity  \cite{goodman,wagner}. The associated density of probability is the Rayleigh law $\rho_R(\lambda)$:
\begin{equation}
\label{eqn:rayleigh_law_detection}
\rho_R(\lambda)=2\lambda \exp \left ( -\lambda^2 \right)
\end{equation}
$\rho_R(\lambda)$ is compared to the experimental estimator $\hat{\rho}_I(\lambda)$ in Fig.\ref{fig:fig10}. The agreement between theory and experiment is excellent. Consequently, the Rayleigh law $\rho_R(\lambda)$ will be considered when a detection criterion is set for the main peak of the echographic image.

Now that the probability density functions of $\tilde{\lambda}_i^0$, $\tilde{\lambda}_i^F$ and $\tilde{I}_l$ are known, a detection criterion can be set for each imaging technique. The relevant quantity for that is the distribution functions $F_1$ of the first singular values, $\tilde{\lambda}_1^F$ and $\tilde{\lambda}_1^0$, and of the main peak of the echographic image. $F_1$ will directly provide the probability of false alarm $PFA$ for the target detection issue, since $PFA(\alpha)=1-F_1(\alpha) = \mbox{Prob} \left \{ \alpha \leq \lambda   \right \}$ (where $\alpha$ is the detection threshold and $\lambda$ is the variable on which the detection threshold is applied).

\begin{figure}[htbp] 
\center
\includegraphics{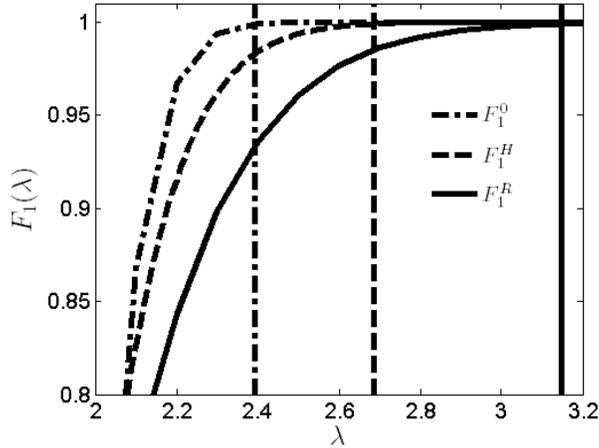}
\caption{\label{fig:fig11}Distribution functions $F^0_1(\lambda)$ (dash-dotted line), $F^H_1(\lambda)$ (dashed line) and $F^R_1(\lambda)$ (continuous line) obtained for $M=32$. The vertical lines represent the detection thresholds $\alpha$ for a PFA $\gamma=10^{-3}$.}
\end{figure}

As we have seen, the distribution $\hat{\rho}_0(\lambda)$ does not strictly follow the quarter-circle law. Thus, the distribution function $F_1^0$ of $\tilde{\lambda}_1^0$ will be estimated from experimental measurements rather than from an analytical expression. By building the histogram of the first singular value $\tilde{\lambda}_1^0$ and then considering its primitive, we obtain an estimator of the distribution function $F_1^0$ of $\tilde{\lambda}_1^0$, which is plotted in Fig.\ref{fig:fig11}. 

In the case of $\mathbf{K^F}$, the distribution of its singular values was found in good agreement with the Hankel law. The distribution $F_1^H$ of the first singular value of a Hankel random matrix is calculated numerically \cite{aubry}. $F_1^H(\lambda)$ is also plotted in Fig.\ref{fig:fig11}.

As to FB, the Rayleigh law $\rho_R(\lambda)$ was found to fit the data properly. Unlike the singular values of a random matrix \cite{mehta,pastur}, the coordinates of the echographic image are independently distributed. In that case, the distribution function $F_1^R(\lambda)$ of the main peak, $\tilde{I}_{\mbox{\small max}}\left(=\mbox{max} \left [ \tilde{I}_l \right] \right)$, of the image can be directly deduced from $\rho_R(\lambda)$. $F_1^R(\lambda)$ is equal to the $M^{th}$ power of the distribution function $F^R(\lambda)$ of any image coordinate:
$$F_1^R(\lambda)= \mbox{Prob} \left \{ \tilde{I}_{\mbox{\small max}} \leq \lambda \right \} = \left [F^R(\lambda) \right]^M\mbox{ ,}$$
$$\mbox{with, }F^R(\lambda)= \int_0^{\lambda} dx \rho_R(x) \mbox{.}$$
The distribution function $F_1^R(\lambda)$ is plotted in Fig.\ref{fig:fig11}.

At this stage, we have reliable models for the distribution function $F_1$ of the relevant variable for all three techniques (DORT method, DORT method combined with the SSF, FB). An admitted probability of error $\gamma$ (\textit{i.e}, a
false alarm rate) is chosen. The three detection thresholds $\alpha$ can be obtained from \cite{aubry}:
\begin{equation}
\label{eqn:detect_thershold}
\alpha = F^{-1}_1(1-\gamma)
\end{equation}
In Fig.\ref{fig:fig11}, the detection thresholds are represented with vertical lines; the admitted PFA $\gamma$ has been set to $10^{-3}$ for all three imaging techniques. The corresponding numerical values are given in Tab.\ref{tab:perf}.

Once the detection thresholds are known, we can also evaluate the performances of each technique for detecting a target. It consists in predicting the signal-to-noise ratio above which the target is detected (``noise'' meaning here multiple scattering). Let $\sigma_T^2$ and $\sigma_M^2$ be the power of signals associated with the target and the multiple scattering contribution. We can predict above which ratio $\frac{\sigma_T}{\sigma_M}$, the target is detected by each technique with the same probability of false alarm $\gamma= 10^{-3}$. Details of calculations are given in Appendix \ref{app:perf}. The performances of each technique are summarized in Tab.\ref{tab:perf}. 
\begin{table}
\caption{\label{tab:perf}Table of detection thresholds deduced from Eq.\ref{eqn:detect_thershold} taking $\gamma=10^{-3}$ and of detection conditions established in Appendix \ref{app:perf}, with $M=32$. }
\begin{center}
\vspace{5mm}
\begin{tabular}{cccc}
\hline
\hline
Imaging technique & DORT, $\mathbf{K^0}$ & DORT, $\mathbf{K^F}$ & FB \\
\hline
Detection threshold & $\alpha=2.39$ & $\alpha=2.69$  &  $\alpha=3.15$ \\
Detection condition & $\,\, \frac{\sigma_T}{\sigma_M} >  \frac{\alpha}{\sqrt{M}}\simeq 0.42\, \,$ & $\, \,\frac{\sigma_T}{\sigma_M} > \frac{\alpha \sqrt{2}}{M}\simeq 0.12\, \,$ & $\,\, \frac{\sigma_T}{\sigma_M} > \frac{\alpha \sqrt{2}}{M} \simeq 0.14\, \,$ \\
\hline
\hline
\end{tabular}
\end{center}
\end{table}

The SSF+DORT approach is the most efficient in terms of detection. Its detection limit decreases in $\frac{1}{M}$ as for FB, which is better by far than the classical DORT method whose detection limit decreases in $\frac{1}{\sqrt{M}}$. For a given probability of false alarm, the SSF+DORT approach succeeds in detecting the target for slightly smaller signal-to-noise ratios than FB (0.12 vs 0.14). As we will see in Sec.\ref{sec:aberration}, this is not the only reason why this approach provides better results than FB: it also diminishes aberration distortions, whereas the image obtained with standard echography can be strongly damaged by aberration.


\begin{figure}[htbp] 
\includegraphics{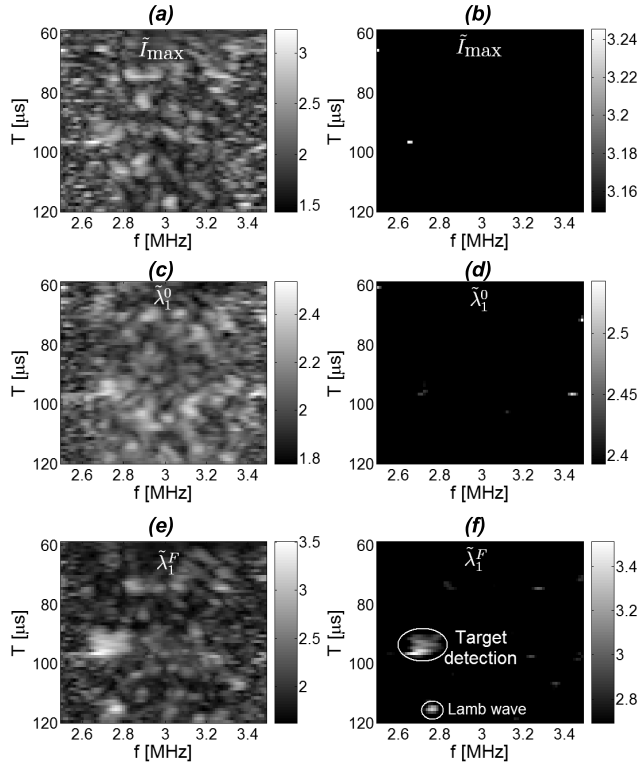}
\caption{
\label{fig:fig12}
Time-frequency evolution of $\tilde{I}_{max}$(top), $\tilde{\lambda}^0_1$(middle), $\tilde{\lambda}^F_1$(bottom). On the left, no criterion of detection is applied. On the right, the gray scales have been adjusted so that for each technique, all values below the detection criterion are represented in black. The probability of false alarm is the same for all techniques ($\gamma=0.1\%$).}
\end{figure}
The detection thresholds summarized in Tab.\ref{tab:perf} are now applied to the experimental results. Fig.\ref{fig:fig12} represents the time-frequency evolution of $\tilde{I}_{max}$, $\tilde{\lambda}_1^0$ and $\tilde{\lambda}_1^F$. From the figures on the left, it is difficult to decide for which time $T$ and frequency $f$, the target is detected. The application of the detection thresholds, based on the same probability of false alarm, provides an unambiguous answer (see Fig.\ref{fig:fig12}). The target is detected over very few time-frequency couples for FB and the classical DORT method. On the contrary, the combination of DORT with the SSF manages to detect the target over the frequency band $2.65-2.8$ MHz and a 7-$\mu$s-long temporal window. It is no accident that the target is best detected around 2.7 MHz. Actually, the forest of rods exhibits a larger scattering mean free path $l_e$ ($\simeq10$ mm) around this frequency \cite{derode4}. Thus, the slab is more transparent in this frequency bandwidth, the direct echo of the target is less attenuated by scattering. Setting detection thresholds with the same PFA provides a systematic way to compare the three techniques and detect the target, which would not have been possible by a simple look at the echographic image (Fig.\ref{fig:echo_temp}) or at singular values (Fig.\ref{fig:fig12}, left column). 

\begin{figure}[htbp] 
\includegraphics{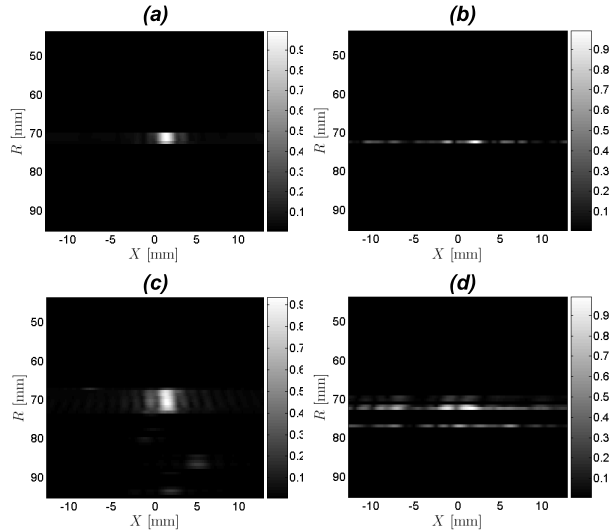}
\caption{\label{fig:fig13}
Images of the target: (a) reference image, without the forest of rods; (b) FB through the forest of rods; (c) DORT combined with the SSF through the forest of rods. (d) DORT alone through the forest of rods. All images have been renormalized by their maximum.}
\end{figure}
\begin{figure}[htbp]
\includegraphics{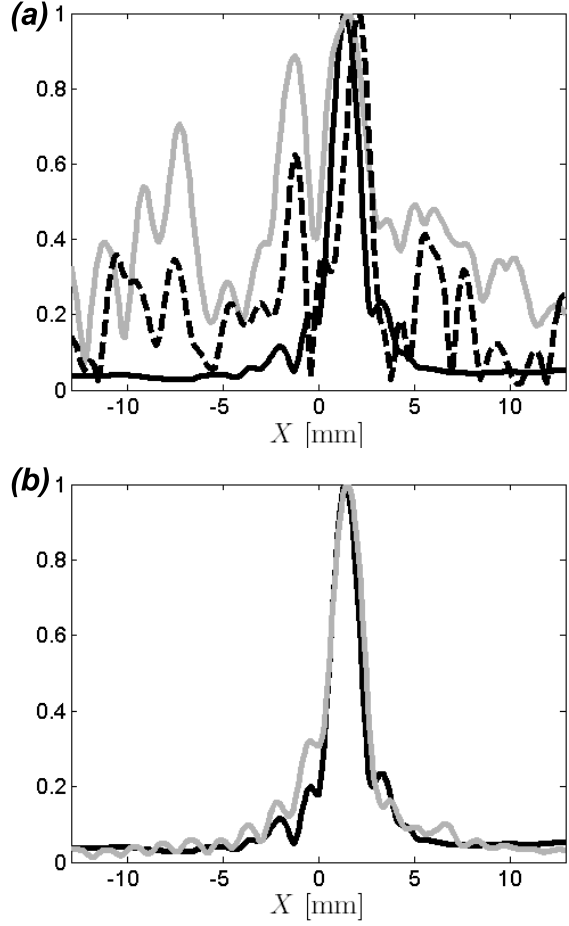}
\caption{\label{fig:fig14}
Section of the images obtained at the depth of the target ($z=68$ mm). The reference 
image without the forest of rods is in continuous black line.  (a) DORT method alone (continuous grey line) and FB (dashed line). (b) DORT method combined with the SSF (continuous grey line). All curves have been renormalized with their maximum.}
\end{figure}

Now that we have determined the time-frequency couples for which the target is detected, the final image can be obtained by summing, for each time $T$, the images over the frequencies $f$ which fulfill the detection criterion. The final image is displayed as a function of the transverse position $X$ and the depth $R=cT/2$. Fig.\ref{fig:fig13} shows the images obtained for each imaging technique. The ``ideal'' image obtained without the forest of rods is also shown and constitutes the reference (Fig.\ref{fig:fig13}(a)). The results are excellent (see Fig.\ref{fig:fig13}(c)): the SSF provides an image of the target which is comparable to the reference image, although its axial resolution is a bit degraded: the temporal spreading of the target echo compared to the ``ideal'' case is due to the loss of a major part of the initial frequency bandwidth. Whereas the emitted signal displays a frequency bandwidth of 1 MHz, the target echo is only detected over a bandwidth of 0.15 MHz. The lateral resolution is nearly as good as the ``ideal'' image (see Fig.\ref{fig:fig14}(b)), and the correct position of the target is obtained.

The other techniques (DORT alone, FB) manage to detect the target, but only in very narrow frequency bands, and they are strongly affected by aberration. Even though one peak is observed around the expected location of the target,  there are secondary lobes (for both techniques) and a displacement of the focal spot (for FB) (see Fig.\ref{fig:fig14}(a)). This is due to the inadequacy of the Born approximation: when backpropagating data, the medium is considered as homogeneous, which is obviously not valid here. The various techniques of aberration corrections \cite{flax,waag,borcea,coucou06,coucou08} are difficult to apply in our experimental configuration, because of multiple scattering. On the contrary, the SSF+DORT approach is less sensitive to the error made when backpropagating data in a supposedly homogeneous medium.  As we will see in the next section, the filtering of the antidiagonals of $\mathbf{K}$ smooths the distortions endured by the wave front, which tends to diminish the aberration effects.

An other interesting observation is the occurrence of an echo above the detection threshold around $T=115 $ $\mu$s (see Fig.\ref{fig:fig12}(f)). This echo is not an artifact due to multiple scattering. It corresponds to circumferential waves that have propagated around the air-filled cylinder. This phenomenon has been already observed with the DORT method \cite{kerbrat}. The difference of arrival times between the specular echo (90 $\mu$s $<T <$97 $\mu$s) and this second echo ($T \simeq 115$ $\mu$s) is compatible with an $A_0$ Lamb mode. This is interesting for a better characterization of the target.

\section{\label{sec:aberration}aberration}

So far, we have dealt with the issue of multiple scattering, and how it could be partially eliminated by a matrix manipulation, in order to improve target detection. But multiple scattering is not the only enemy in imaging and detection. Even if only single scattering takes place, a heterogeneous layer such as the forest of rods induces aberrations that distort the wavefront reflected by the target. To reconstruct an image, both FB and DORT rely on the hypothesis that the medium has a constant speed of sound, which is clearly not true. In this section, we examine the impact of the SSF on aberrations. We consider the same experimental set-up, except that now the aberration undergone by the target echo will be examined independently from the multiple scattering contribution of the forest of rods. To that end, the impulse response matrix has been measured in three configurations:
\begin{itemize}
\item Configuration 1: With the scattering slab and the target; it corresponds to the experimental situation studied until now.
\item Configuration 2: With the scattering slab alone (the target has been removed).
\item Configuration 3: With the target alone (the scattering slab has been removed).
\end{itemize}

\begin{figure}[htbp] 
\includegraphics{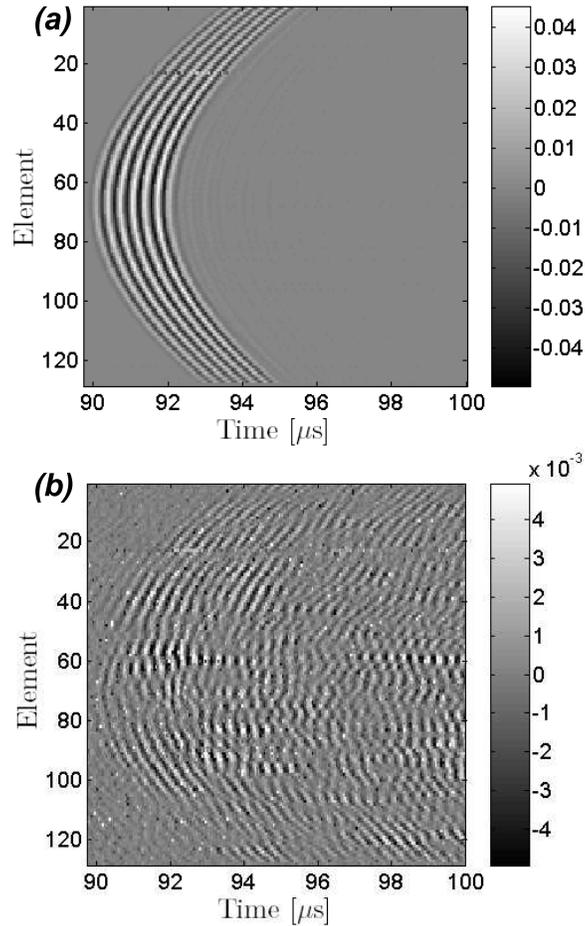}
\caption{\label{fig:fig15}(a) Line 64 of matrix $\mathbf{H}^{(3)}(t)$ in the time-window $ 90-100$ $\mu$s. (b) Line 64 of matrix $\mathbf{H}(t)$ in the same time-window.}
\end{figure}

Let $\mathbf{H}^{(i)}(t)$, denote the corresponding matrices, where the superscript $i$ stands for the configuration (1, 2 or 3). In order to investigate aberration effects apart form multiple scattering, we calculate the matrix $\mathbf{H}=\mathbf{H}^{(1)}-\mathbf{H}^{(2)}$. $\mathbf{H}$ contains only signals linked to the target. Particularly, its first arrivals correspond to the single scattering contribution (ballistic) coming from the target. The later echoes correspond to multiple scattering paths involving both the target and the forest of rods. In Fig.\ref{fig:fig15}, a line of the matrix $\mathbf{H}(t)$ is displayed as a function of the arrival time $t$ in the time window $90-100$ $\mu$s. It is compared with the same line of matrix $\mathbf{H}^{(3)}(t)$. The comparison of matrices $\mathbf{H}$ and $\mathbf{H}^{(3)}$ highlights the phase and amplitude distortions undergone by the wave-front reflected by the target when it comes through the scattering slab.

\begin{figure}[htbp] 
\includegraphics{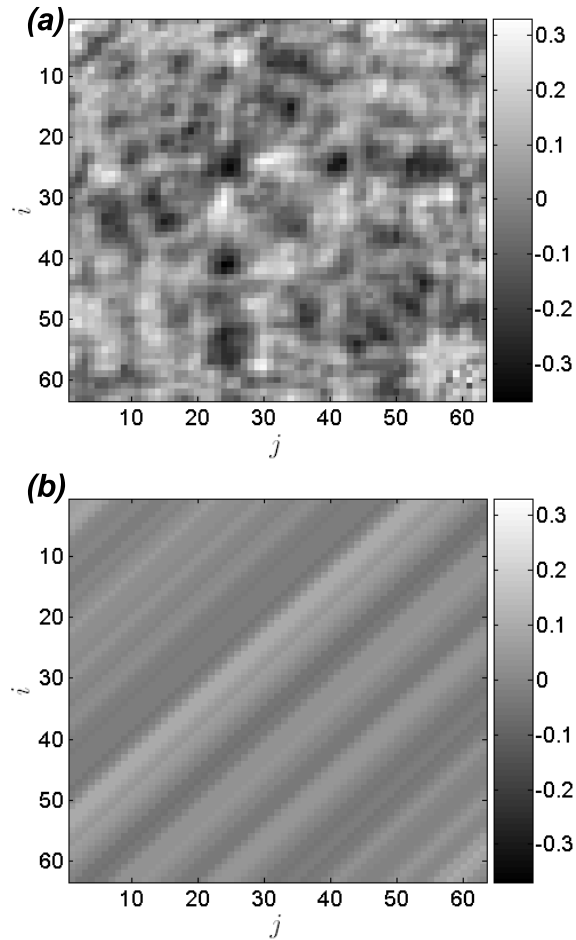}
\caption{\label{fig:fig16}(a) Real part of the distortion matrix $\mathbf{D^0}$ at the frequency $f=3.1$ MHz. (b) Real part of the filtered distortion matrix $\mathbf{D^F}$ at the same frequency.}
\end{figure}

Let us study the action of the SSF on the distortions. The frequency spectrum of $\mathbf{H}$ and $\mathbf{H}^{(3)}$ is calculated by means of a DFT in the time-window $ 90-100$ $\mu$s. Two matrices, $\mathbf{K}(f)$ and $\mathbf{K}^{(3)}(f)$, are obtained at each frequency $f$. The distortions induced by the scattering slab can be quantified by a matrix $\mathbf{D}(f)$ whose coefficients $d_{ij}(f)$ are
\begin{equation}
\label{eqn:matrixA}
d _{ij}(f)=\frac{k_{ij}(f)}{k^{(3)}_{ij}(f)}
\end{equation}
If the scattering slab had no effect, the coeffients $d_{ij}$ would be real (no phase distortion) and equal to unity (no amplitude aberration). Obviously, in our case, the coefficients $d_{ij}$ show both amplitude and phase distortions (Fig.\ref{fig:fig16}).

Let us express the coefficients $k_{ij}$ and $k^{(3)}_{ij}$. Considering Eq.\ref{eqn:eq_ss_signal} with only one scatterer (the target itself) we have:
\begin{equation}
\label{eqn:eq_target}
k_{ij}^{(3)}(f) \propto \exp \left [ j k \frac{ \left (x_i -X_T \right )^2}{2R_T} \right ] \exp \left [ j k \frac{ \left (x_j -X_T \right )^2}{2R_T} \right ]
\end{equation}
where the coordinates $(X_T, R_T)$ correspond to the target location. For the sake of simplicity, we have removed the phase term $\frac{\exp \left ( j 2kR_T \right )}{R_T}$ and the reflectivity term in Eq.\ref{eqn:eq_ss_signal}, which are unimportant here. Using Eq.\ref{eqn:matrixA} and Eq.\ref{eqn:eq_target}, we have
\begin{equation}
\label{eqn:eq_target_aberration}
k_{ij}(f) \propto  d _{ij}(f) \exp \left [ j k \frac{ \left (x_i -X_T \right )^2}{2R_T} \right ] \exp \left [ j k \frac{ \left (x_j -X_T \right )^2}{2R_T} \right ]
\end{equation}

As seen in Sec.\ref{sec:anti_filtering}, the SSF consists in projecting the antidiagonals of the matrix $\mathbf{K}$ on the characteristic space of the single scattering contribution. Once this operation is performed, the coefficients of the filtered matrix $\mathbf{K^F}$ can be expressed as (see Appendix \ref{sec:app_aberration})
\begin{equation}
\label{eqn:eq_target_aberration_SSfilter}
k_{lm}^{F}(f) \propto  e_{l+m-1}(f) \exp \left [ j k \frac{ \left (x_l -X_T \right )^2}{2R_T} \right ] \exp \left [ j k \frac{ \left (x_m -X_T \right )^2}{2R_T} \right ]
\end{equation}
where the coefficients $e_v$ are given by
\begin{eqnarray}
& & \mbox{if  }   v  \, \,\mbox{is an odd number,} \nonumber\\
& & \mbox{then,  }  e_{v}=  \left < d \left [u+\frac{v-1}{2}, \frac{v-1}{2}-u+2M-1 \right ] \right >_{u=1,...,2M-1} \\
& & \mbox{if  }   v \,\,\mbox{is an even number,} \nonumber \\
& & \mbox{then,  }e_{v}=  \left < d \left [u+\frac{v}{2},\frac{v}{2}-u+2M-1 \right ] \right >_{u=1,...,2M-2}
\end{eqnarray}
and the symbol $<.>$ denotes an average over the variable in the subscript. Therefore the coefficients $e_v$ result from a smoothing of $d_{ij}$: the effect of the SSF is to average the distortion coefficients along each antidiagonal. Fig.\ref{fig:fig16}(b) represents the filtered distortion matrix, $\mathbf{D^F}$. Its coefficients are $d_{ij}^F=e_{i+j-1}$. The comparison of matrices $\mathbf{D^0}$ and $\mathbf{D^F}$ shows that the SSF reduces the fluctuations of the distortion coefficients (see Fig.\ref{fig:fig16}).

\begin{figure}[htbp] 
\includegraphics{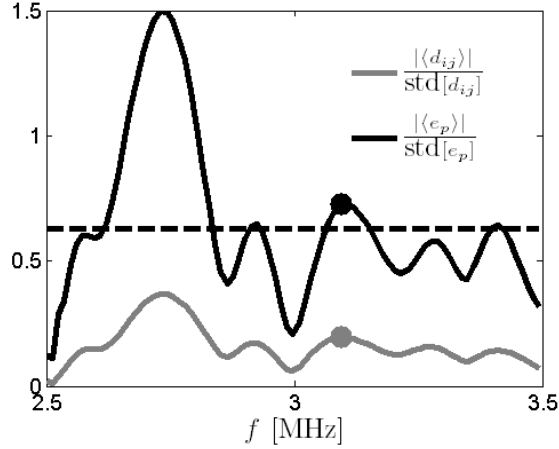}
\caption{\label{fig:fig17}Ratios $\left | \left < d_{ij} \right>\right | /\mbox{std}\left [d_{ij} \right ] $ (continous grey line) and $\left | \left < e_p \right>\right | /\mbox{std}\left [e_p\right ] $(black continous line) are plotted as a function of frequency. The horizontal dashed line corresponds to the detection threshold of Eq.\ref{eqn:lambda_1}}
\end{figure}

The standard deviation $\mbox{std}\left [e_p \right ]$ of coefficients $e_p$ is smaller than $\mbox{std}\left [d_{ij} \right ]$ by a factor of $\sqrt{N_{ind}}$. 
$N_{ind}$ is the number of independent elements along each antidiagonal of matrix $\mathbf{D}$, which depends on the coherence length of the aberrator relatively to the array pitch. If the matrix $\mathbf{D}$ contained independent entries, $N_{ind}$ would be equal to $M$. In our experimental configuration, there are correlations between neighbouring elements belonging to the same antidiagonal, and $N_{ind} \sim \frac{M}{2}$. Fig.\ref{fig:fig17} illustrates the action of the SSF on the fluctuations of the distortion coefficients. The evolution of the ratios $\left | \left < d_{ij} \right > \right |/\mbox{std}\left [d_{ij} \right ]$ and $\left | \left < e_p \right > \right |/\mbox{std}\left [e_{p} \right ]$ is plotted as a function of frequency. Note that $\left | \left < e_p \right > \right |\equiv \left | \left < d_{ij} \right > \right |$. The ratio between both curves is close to $\sqrt{N_{ind}}\simeq \sqrt{\frac{M}{2}}=4$. 

Now that it is clear that the SSF has a smoothing effect on the aberrations, let us evaluate by how much it will improve target detection. We start by keeping only one in four elements of the filtered matrix $\mathbf{K^F}$: its new dimensions are $\frac{M}{2} \times \frac{M}{2}$.
As before (see Sec.\ref{sec:detect_crit}), this is done to remove short-range correlations between matrix elements, which allows us to use relatively simple results derived from RMT. Next, we perform a singular value decomposition:
\begin{equation}
\label{eqn:dort2}
\mathbf{K^F} = \mathbf{U^F} \mathbf{\Lambda^F} \mathbf{V^{F\dag}}
\end{equation}
If there were no aberrations at all, backpropagating as usual the first singular vector $\mathbf{V_1^F}$ would focus at the target position. This is not the case here, because of aberrations induced by the scattering layer. In order to analyse their effect, let us write $\mathbf{K^F}$ as the sum of a ``smoothed'' matrix $\left< \mathbf{K^F} \right > $ and a perturbation $\mathbf{\Delta K^F}$:
\begin{equation}
\label{eqn:perturb_aberration}
\mathbf{K^F} =  \underbrace { \left< \mathbf{K^F} \right > } _{\mbox{Matrix of rank 1}}+ \underbrace {\mathbf{\Delta K^F}}_{\mbox{Random Hankel matrix}}
\end{equation}
Note that the absence of subscript behind the symbol $<.>$ means that we now consider ensemble averages. From Eq.\ref{eqn:eq_target_aberration_SSfilter}, we have the coefficients of $\left< \mathbf{K^F} \right >$ and $\mathbf{\Delta K^F}$:
\begin{equation}
\label{eqn:perturb_aberration2}
\left< k^{F}_{lm}(f)\right >= \left < e_{l+m-1}(f) \right > \exp \left [ j k \frac{ \left (x_l -X_T \right )^2}{2R_T} \right ] \exp \left [ j k \frac{ \left (x_m -X_T \right )^2}{2R_T} \right ]
\end{equation}
\begin{equation}
\label{eqn:perturb_aberration3}
\delta k^{F}_{lm}(f) = \left [  e_{l+m-1}(f)- \left < e_{l+m-1}(f) \right > \right ]  \exp \left [ j k \frac{ \left (x_l -X_T \right )^2}{2R_T} \right ] \exp \left [ j k \frac{ \left (x_m -X_T \right )^2}{2R_T} \right ]
\end{equation}
Note that $\left< \mathbf{K^F} \right >$ may be written as
\begin{equation}
\label{eqn:perturb_aberration_4}
\left < \mathbf{K^F} \right > = \left <e_p \right > \mathbf{K}^{(3)} 
\end{equation}
with $\left <e_p \right >$ the average distortion and $ \mathbf{K}^{(3)} $ the response of the target without the scattering slab.
$\left< \mathbf{K^F} \right >$ is of rank 1 and its only singular vector focuses at the exact location of the target. $\mathbf{\Delta K^F}$ is the perturbation due to the aberration effects: it explicitly depends on the fluctuations of the distortion coefficients $e_p$ around their mean $\left <e_p \right >$. $\mathbf{\Delta K^F}$ has a particular feature: since the coefficients $\left [  e_{l+m-1}(f)- \left < e_{l+m-1}(f) \right > \right ]$ are constant along each antidiagonal ($l + m$ =constant), there is a deterministic phase relation between the coefficients of $\mathbf{\Delta K^F}$ located on the same antidiagonal. As shown in \cite{aubry}, this kind of matrix has the same statistical behavior as a random Hankel matrix.
Backpropagating the first singular vector of $\mathbf{K^F}$ will focus at the target position, as long as the perturbation $\mathbf{\Delta K^F}$ is weak compared to $\left< \mathbf{K^F} \right >$. A threshold can be evaluated, based on RMT (the details are given in Appendix \ref{sec:app_aberration_2}). The SVD will successfully extract $\left< \mathbf{K^F} \right >$ as long as
\begin{equation}
\label{eqn:lambda_1}
\frac{\left | \left < e_p \right > \right |}{\mbox{std}\left [ e_p \right ]} > \frac{\alpha} {\sqrt{M/2}}
\end{equation}
with $\alpha$ the detection threshold found for the first normalized singular value in the case of a random Hankel matrix (see Sec.\ref{sec:detect_crit}). This detection threshold is displayed with a horizontal line in Fig.\ref{fig:fig17}. $\alpha$ has been calculated here considering the distribution function $F_1^H(\lambda)$ obtained for a random Hankel matrix of size $\frac{M}{2} \times \frac{M}{2}$, with $\frac{M}{2}$ =16. As before, the PFA has been fixed to $\gamma=10^{-3}$ and we have obtained numerically the threshold $\alpha=2.52$.  The SSF+DORT approach succeeds at the frequencies $f$ where the ratio $\left | \left < e_p \right > \right | / \mbox{std}\left [ e_p \right ]$ (black curve) is above the detection threshold defined in Eq.\ref{eqn:lambda_1} (see Fig.\ref{fig:fig17}).

\begin{figure}[htbp] 
\includegraphics{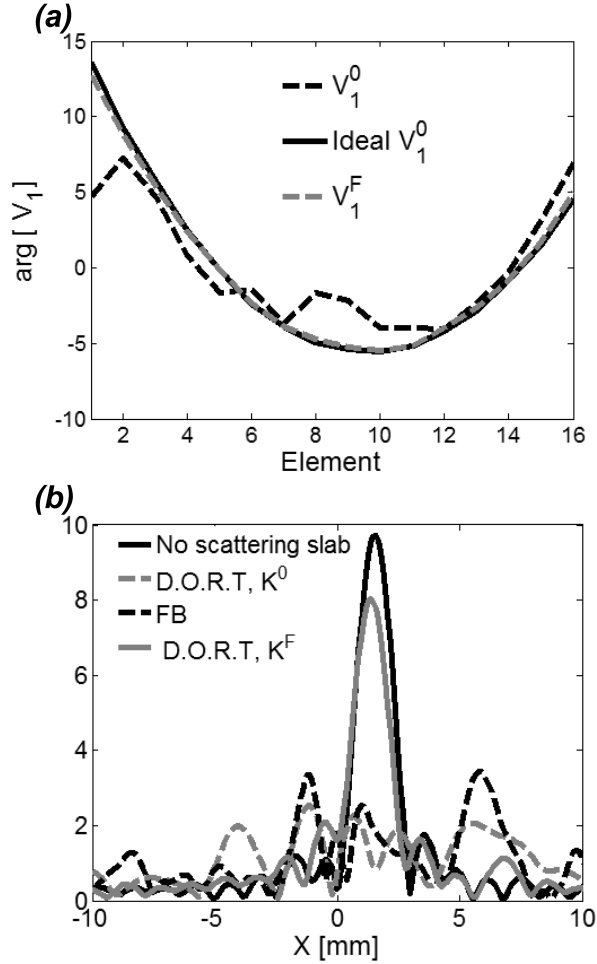}
\caption{\label{fig:fig18} 
Influence of the phase distortions on the three imaging techniques at frequency $f$ = 3.1 MHz (a) Unwrapped phase of the first singular vectors $\mathbf{V^0_1}$(black dashed line) and $ \mathbf{V^F_1}$ (gray dashed line) compared with the ``ideal'' wave-front (continuous black line) which would be obtained if the slab was removed. (b) Images of the target obtained with FB (black dashed line), DORT applied to $\mathbf{K^0}$ (grey dashed line) and DORT applied to $\mathbf{K^F}$ (grey continuous line). The ideal image obtained without the scattering slab is in continuous black line. All curves have been renormalized with their mean
amplitude.}
\end{figure}

As an illustration, we choose to work at $f=3.1$ MHz, a frequency for which we are just above the threshold of Eq.\ref{eqn:lambda_1}. The SVD is applied to the matrices $\mathbf{K^0}$ and $\mathbf{K^F}$. The unwrapped phases of the first singular vectors $ \mathbf{V^0_1}$ and $ \mathbf{V^F_1}$ are plotted in Fig.\ref{fig:fig18}(a). They are compared with the ``ideal'' phase which is obtained without the scattering slab. It is given by the parabolic term $ k \frac{ \left (x_i -X_T \right )^2}{2R_T} $ which allows to focus on the target when the first singular vector is backpropagated numerically. The unwrapped phase of the first singular vector is a relevant observable because it directly shows the phase distortions of the wave front which focuses on the target.
Without prior filtering, the strong phase distortions of the wave-front result in a first singular vector $ \mathbf{V^0_1}$ whose unwrapped phase exhibits erratic deviations from the ``ideal'' parabolic law. On the contrary, the SSF leads to a first singular vector $ \mathbf{V^F_1}$ whose unwrapped phase is very close to the ``ideal'' case (Fig.\ref{fig:fig18}(a)). 

Fig.\ref{fig:fig18}(b) displays the images obtained with the numerical backpropagation of the singular vectors on the focal plane. The backpropagation of $\mathbf{V^F_1}$ focuses nicely on the target, nearly as if there was no aberrating slab. On the contrary, without prior filtering, the singular vector $\mathbf{V^0_1}$ does not focus on the target and it is impossible to deduce the target location from the image. Finally, we also show the result of FB applied to the matrix  $\mathbf{K^0}$ (see Sec.\ref{subsec:frequential_ultrasono}). Once again, the phase distortions induced by aberration are so large that the image displays several main lobes with no connection with the target location. 

In this example, the action of the SSF on phase distortions is obvious.
The fluctuations of the distortion coefficients $d_{ij}$ are diminished by a factor of $\sqrt{\frac{M}{2}}$. As long the criterion (Eq.\ref{eqn:lambda_1}) is fulfilled, the SVD succeeds in extracting the unaberrated part $\left < \mathbf{K^F} \right >$ from the measured matrix. Then, backpropagting the first singular vector $\mathbf{V^F_1}$ provides the correct target location as shown in Fig.\ref{fig:fig18}(b).

In the previous sections, we have compared the SSF to FB and showed that it gave much better results when trying to detect a target behind a multiple scattering and aberrating layer. The smoothing of the wave-front distortions provided by the SSF partly accounts for this. Even if the single scattered echo coming from the target is sufficiently large compared to the multiple scattering contribution, FB may fail in detecting the target because of the strong phase distortions induced by the scattering slab. On the contrary, the SSF smoothes the phase distortions enough for the SVD to extract the undistorted wave-front. Backpropagating the first singular vector in a virtually homogeneous medium finally allows to image the target, with no secondary lobes, and no displacement of the focal spot.

\section{\label{sec:conclusion}Conclusion}

The approach we developed here combines a ``single scattering filter'' with the DORT method. It greatly improves the performance of an array of transmitters/receivers in detecting and imaging a target hidden behind a scattering medium. On the one hand, the short time-frequency analysis allows to select the frequency bandwidth(s) favourable to the detection of the target, unlike classical echography which is performed in the temporal domain. This has been made possible by setting a detection criterion based on random matrix theory. We showed that removing most of the multiple scattering contribution significantly improves the performance of the DORT method in random scattering media. The results are even better than focused beamforming in terms of target detection. On the other hand, the SSF+DORT approach is shown to strongly diminish the influence of aberration effects (secondary lobes, displacement of the focal spot) which burden classical imaging techniques. The perspectives of this study are numerous. This technique can be applied to other types of waves (electromagnetic, seismic, \textit{etc.}) as long as a coherent array of independent elements is available. A future step will be to test this approach in real situations, such as the detection of a target embedded in the soil or of cracks in concrete structures, steel blocks \textit{etc.}

\section{Acknowledgments}
The authors wish to acknowledge the groupe de recherches IMCODE of CNRS (GDR 2253).

\appendix

\section{\label{app:non_complement}FB applied to $\mathbf{K^F}$}
The aim of this appendix is to show why FB and the SSF are not complementary. 

Using Eq.\ref{eqn:echo_intensity} and under the paraxial approximation, the coordinate $I_l$ of the echographic image at the arrival time $T=2R/c$ and frequency $f$ can be expressed as
\begin{equation}
\label{eqn:appendix_1}
I_l(T,f)= \left | \sum_i \sum_j k_{ij}(T,f) \exp \left \{ -j \frac{k}{2R} \left [\left( x_i-x_l \right)^2 + \left( x_j-x_l \right)^2\right] \right \}\right |
\end{equation}
One can write Eq.\ref{eqn:appendix_1} in the basis defined in Eq.\ref{eqn:sub_var},
\begin{equation}
\label{eqn:appendix_2}
I_l(T,f)=\left | \sum_u \sum_v a_{uv} \exp \left \{ -j \frac{k}{2R} \left [y_u^2+\left( y_v-\sqrt{2}x_l \right)^2\right] \right \} \right| 
\end{equation}
where the coefficients $a_{uv}$ are deduced from $k_{ij}$ after the data rotation described in Sec.\ref{subsec:rot_anti}. Eq.\ref{eqn:appendix_2} can be simplified into 
\begin{equation}
\label{eqn:appendix_3}
I_l(T,f)=\Bigl | \sqrt{L}\sum_v  \exp \left \{ -j \frac{k}{2R} \left [\left( y_v-\sqrt{2}x_l \right)^2\right] \right \} \underbrace{\sum_u s_u^* a_{uv}}_{p_v}  \Bigr |
\end{equation}
where the coordinates $s_u$ of the vector $\mathbf{S}$ are given by Eq.\ref{eqn:vector_s}. We see here that the principle of FB can be decomposed into two steps corresponding respectively to the sum over index $v$ and the sum over index $u$. Actually, the second sum corresponds to the projection of the columns of matrix $\mathbf{A}$ on the characteristic space of single scattering generated by the vector $\mathbf{S}$, as seen in Sec.\ref{subsec:filter_anti}. Thus, the SSF constitutes one of the two steps of FB.

Now, one can try to combine the SSF with FB. We can express the image $I_l^F(T,f)$ that we would obtain from the filtered matrix $\mathbf{K^F}$. It can be deduced from Eq.\ref{eqn:appendix_3}, replacing $a_{uv}$ by the filtered elements $a^F_{uv}$,
\begin{equation}
\label{eqn:appendix_4}
I^F_l(T,f)=\left | \sqrt{L}\sum_v  \exp \left \{ -j \frac{k}{2R} \left [\left( y_v-\sqrt{2}x_l \right)^2\right] \right \} \sum_u s_u^* a^F_{uv} \right |
\end{equation}
Using the expression of $a^F_{uv}$ given in Eq.\ref{eqn:Af_coord_2}, $I^F_l(T,f)$ becomes:
\begin{eqnarray*}
I^F_l(T,f)& =& \left | \sqrt{L}\sum_v  \exp \left \{ -j \frac{k}{2R} \left [\left( y_v-\sqrt{2}x_l \right)^2\right] \right \} \sum_u s_u^* a^S_{uv} \right. \\
&+& \left. \sqrt{L}\sum_v  \exp \left \{ -j \frac{k}{2R} \left [\left( y_v-\sqrt{2}x_l \right)^2\right] \right \} \sum_u s_u^* s_u \sum_{u'=1}^L s_{u'}^*a^M_{u'v} \right |
\end{eqnarray*}
As the vector $\mathbf{S}$ is normalized, we have $\sum_u s_u^* s_u=1$ and the latter equation can be simplified into
\begin{eqnarray}
\label{eqn:appendix_5}
I^F_l(T,f)& =& \left | \sqrt{L}\sum_v  \exp \left \{ -j \frac{k}{2R} \left [\left( y_v-\sqrt{2}x_l \right)^2\right] \right \} \sum_u s_u^* \left [ a^S_{uv}+a^M_{uv}\right] \right | \\
&=& \left | \sqrt{L}\sum_v  \exp \left \{ -j \frac{k}{2R} \left [\left( y_v-\sqrt{2}x_l \right)^2\right] \right \} \sum_u s_u^*a_{uv} \right |
\end{eqnarray}
This equation is strictly identical to Eq.\ref{eqn:appendix_3}:
\begin{equation}
\label{eqn:appendix_6}
I_l(T,f)=I^F_l(T,f)
\end{equation}
which means that the images built from the raw data (matrix $\mathbf{K}$) or the filtered data (matrix $\mathbf{K^F}$) are identical. There is no interest in combining the SSF with FB.

\section{\label{app:perf}Detection condition in presence of multiple scattering}

The aim of this appendix is to predict the performances of each technique (FB, DORT alone, DORT combined with the SSF) in detecting a target hidden behind a diffusive slab. Let $\sigma_T^2$ and $\sigma_M^2$ be the power of signals associated with the target and the multiple scattering contribution. The performance of each technique is assessed by determining the signal-to-noise ratio $\frac{\sigma_T}{\sigma_M}$ above which the target will be detected.

We consider a time of flight $T$ corresponding to the arrival time for the target echo. The measured matrix $\mathbf{K^0}$ (of dimension $M\times M$) can be decomposed as:
\begin{equation}
\label{eqn:K0_decompos_detection}
\mathbf{K^0}=\mathbf{K^T}+\mathbf{K^M}
\end{equation}
$\mathbf{K^M}$ corresponds to the multiple scattering contribution; its coefficients are assumed to be gaussian complex random variables, identically and independently distributed, of variance $\sigma_M^2$ and with zero mean. Because of spatial reciprocity, $\mathbf{K^M}$ is symmetric. $\mathbf{K^T}$ is associated with the target echo. Its coefficients can be expressed as :
\begin{equation}
\label{eqn:eq_target_app_detection}
k_{ij}^{T} = \sigma_T \exp \left [ j k \frac{ \left (x_i -X_T \right )^2}{2R_T} \right ] \exp \left [ j k \frac{ \left (x_j -X_T\right )^2}{2R_T} \right ]
\end{equation}
where $(X_T, R_T)$ are the coordinates of the target. The aberration effects generated by the diffusive slab are neglected.

\subsection{Focused beamforming}
The coordinates of the image $\mathbf{I}$ are given by :
\begin{equation}
\label{eqn:appendix2_1_detection}
I_l=\left | \sum_i \sum_j k^0_{ij} \exp \left \{ -j \frac{k}{2R_T} \left [\left( x_i-x_l \right)^2 + \left( x_j-x_l \right)^2\right] \right \} \right|
\end{equation}

Let us express the intensity of the image, using the decomposition $k_{ij}^0=k_{ij}^T+k_{ij}^M$:
\begin{gather}
I_l^2 =  \underbrace {\sum_{p,q,r,s} k^T_{pq} k^{T*}_{rs} \exp \left \{ -\frac{jk}{2R_T} \left [\left( x_p-x_l \right)^2 - \left ( x_r-x_l \right)^2 + \left( x_s-x_l \right)^2 - \left( x_t-x_l \right)^2\right] \right \}}_{\left [ I_l^{T} \right ]^2} \label{eqn:appendix3_1_detection}\\
+  \underbrace {\sum_{p,q,r,s} k^M_{pq} k^{M*}_{rs} \exp \left \{ - \frac{jk}{2R_T} \left [\left( x_p-x_l \right)^2 - \left ( x_r-x_l \right)^2 + \left( x_s-x_l \right)^2 - \left( x_t-x_l \right)^2\right] \right \}}_{\left [ I_l^{M} \right ]^2} \label{eqn:appendix4_1_detection} \\
+  \underbrace{ \sum_{p,q,r,s}   k^T_{pq} k^{M*}_{rs} \exp \left \{ - \frac{jk}{2R_T} \left [\left( x_p-x_l \right)^2 - \left ( x_r-x_l \right)^2 + \left( x_s-x_l \right)^2 - \left( x_t-x_l \right)^2\right] \right \}}_{ I_l^{T} I_l^{M*}} \label{eqn:appendix5_1_detection} \\
+  \underbrace{ \sum_{p,q,r,s}   k^M_{pq} k^{T*}_{rs} \exp \left \{ - \frac{jk}{2R_T} \left [\left( x_p-x_l \right)^2 - \left ( x_r-x_l \right)^2 + \left( x_s-x_l \right)^2 - \left( x_t-x_l \right)^2\right] \right \}}_{   I_l^{M} I_l^{T*}} \label{eqn:appendix5b_1_detection}
\end{gather}
By injecting the expression of $k^T_{ij}$ into $\left [ I_l^{T} \right ]^2$ (Eq.\ref{eqn:appendix3_1_detection}), the intensity of the target peak is:
\begin{equation}
\left [ I_l^{T} \right ]^2=M^4 \sigma_T^2 \delta(x_l-X_T)
\end{equation} 
where $\delta$ denotes the Kronecker symbol. The multiple scattering contribution (Eq.\ref{eqn:appendix4_1_detection}) results in an image of speckle whose mean intensity $\left < \left [I_l^{M} \right]^2\right >$ is given by:
\begin{equation}
\left <\left [ I_l^{M} \right ]^2 \right > =2 M^2 \sigma_M^2
\end{equation} 
The factor 2 comes from the fact that $k_{ij}^M=k_{ji}^M$.
The third and fourth terms, $I_l^{T} I_l^{M*}$ and $I_l^{M} I_l^{T*}$, correspond to the interference between the signals associated with the target and the multiple scattering contribution. These signals are totally decorrelated, hence
\begin{equation}
\left < I_l^{T} I_l^{M*} \right > =\left < I_l^{M} I_l^{T*}\right >= 0 \\
\end{equation}
In average, the intensity of the echographic image, at the target arrival time, exhibits the following profile: a peak linked to the target at $x_l=X_T$, of intensity $ M^4 \sigma_T^2 $, buried in a speckle pattern whose mean intensity is $2 M^2 \sigma_M^2$. 

If the maximum of the image, $I_{\mbox{\small max}}$, is actually linked to the target, then its amplitude is given by:
\begin{equation}
I_{\mbox{\small max}} \simeq \mbox{E} \left [ I_{\mbox{\small max}} \right] = \sqrt{M^4 \sigma_T^2 + 2 M^2 \sigma_M^2}
\end{equation}
The quadratic mean of the image is:
\begin{equation}
\sqrt{\frac{1}{M}\sum_{l=1}^M I_l^2} \simeq \sqrt{M^3 \left ( \sigma_T^2 + 2\sigma_M^2 \right)}
\end{equation}
Upon normalization (Eq.\ref{eqn:renorm_coord_detection}), $\tilde{I}_{\mbox{\small max}}$ is thus given by:
\begin{equation}
\label{eqn:Imax_appendixB}
\tilde{I}_{\mbox{\small max}} \simeq \frac{\sqrt{M^4 \sigma_T^2 + 2 M^2 \sigma_M^2}}{\sqrt{M^2 \left ( \sigma_T^2 + 2\sigma_M^2 \right)}}
\end{equation}
At the limit of detection, we can assume that $M^2\sigma_T^2 >> \sigma_M^2 >>\sigma_T^2$. The validity of this approximation will be proved \textit{a posteriori} by the final result. Eq.\ref{eqn:Imax_appendixB} simplifies into:
\begin{equation}
\tilde{I}_{\mbox{\small max}} \simeq \frac{M\sigma_T}{\sqrt{2} \sigma_M}
\end{equation}
As to FB, the detection threshold corresponds to the condition $\tilde{I}_{\mbox{max}}>\alpha$, where $\alpha$ is given by Eq.\ref{eqn:detect_thershold} and depends on the PFA $\gamma$. Finally, we obtain the following detection criterion:
 \begin{equation}
 \label{eqn:appendix6_1_detection}
\frac{\sigma_T}{\sigma_M}>\frac{\alpha \sqrt{2}}{M}
\end{equation}
This condition is reported in Tab.\ref{tab:perf}. It indicates the signal-to-noise ratio $\frac{\sigma_T}{\sigma_M}$ above which the main peak will correspond to the target (with a probability of false alarm $\gamma$). Note that Eq.\ref{eqn:appendix6_1_detection} is only valid for a multiple scattering noise, which is spatially reciprocal. If we had dealt with an additional noise which does not respect this property, the detection criterion would be:
$$\frac{\sigma_T}{\sigma_M}>\frac{\alpha}{M}$$

\subsection{\label{subsec:app2_dort}The DORT method}
First, we deal with the classical DORT method, \textit{i.e} when the SVD is applied directly to the raw matrix $\mathbf{K^0}$. If the first singular value $\lambda_1^0$ is actually linked to the target echo, then its expected value is given by \cite{aubry}:
$$\mbox{E}\left [ \lambda^0_1 \right ]= M \sigma_T$$
The quadratic mean of singular values is given by \cite{aubry}:
$$  \sqrt { \frac{1}{M}\sum_{p=1} \left [\lambda^0_p\right ]^2}\simeq\sqrt{M\left ( \sigma_T^2+\sigma_M^2 \right)} $$ 
Upon normalization(Eq.\ref{eqn:eq2}),  the expected value of $\tilde{\lambda}^0_1$ is thus given by:
$$\mbox{E} \left \{ \tilde{\lambda}^0_1 \right \}=\sqrt{M\frac{\sigma_T^2}{\sigma_T^2+\sigma_M^2}}\simeq \frac{\sigma_T}{\sigma_M}\sqrt{M} \,\,\mbox{,
for} \,\, \sigma_T^2<<\sigma_M^2$$
The application of the detection criterion $\tilde{\lambda}^0_1 > \alpha$ leads to the following detection condition:
\begin{equation}
\label{eqn:cond_detection_K0}
\frac{\sigma_T}{\sigma_M} > \frac{\alpha}{\sqrt{M}}
\end{equation}
This condition is reported in Tab.\ref{tab:perf}. If we compare it with the one obtained above for FB (Eq.\ref{eqn:appendix6_1_detection}), we see that the DORT method is clearly more sensitive to noise than FB. The detection criterion varies as $M^{-1/2}$ for the DORT method, whereas it is $M^{-1}$ for FB.

The argument is the same for the DORT method applied to $\mathbf{K^F}$, except that we have to take into account the action of the SSF. In Sec.\ref{subsec:filter_anti}, we have shown that the filtering of antidiagonals decreases the multiple scattering contribution by a factor $\sqrt{\frac{M}{2}}$. Thus, the filtered matrix $\mathbf{K^F}$ can be decomposed as:
\begin{equation}
\label{eqn:KF_decompos_detection}
\mathbf{K^F}=\mathbf{K^T}+\mathbf{K^{MF}}
\end{equation}
The matrix $\mathbf{K^{MF}}$ is linked with the residual contribution of multiple scattering. $\mathbf{K^{MF}}$ is a random Hankel matrix whose coefficients have a variance of $2\sigma_M^2/M$. The results obtained for $\mathbf{K^0}$ can be applied directly to $\mathbf{K^F}$, taking into account the lower variance of multiple scattering signals. If the first singular value $\lambda_1^F$ is associated to the target echo, then its expected value is still given by:
$$\mbox{E}\left [ \lambda^F_1 \right ]= M \sigma_T$$
The quadratic mean of singular values is given by :
$$  \sqrt { \frac{1}{M}\sum_{p=1} \left [\lambda^F_p\right ]^2}=\sqrt{M \left ( \sigma_T^2+\frac{2\sigma_M^2}{M} \right)} $$ 
Finally, upon normalization (Eq.\ref{eqn:eq2}), the expected value of  $\tilde{\lambda}^F_1$ is given by:
$$\mbox{E} \left \{ \tilde{\lambda}^F_1 \right \}=\sqrt{M \frac{\sigma_T^2}{\sigma_T^2+\frac{2\sigma_M^2}{M}}}\simeq \frac{M}{\sqrt{2}}\frac{\sigma_T}{\sigma_M} \,\,\mbox{,
for} \,\, M\sigma_T^2<<\sigma_M^2$$
The validity of the approximation $ M \sigma_T^2<< \sigma_M^2$ will be proved \textit{a posteriori} by the final result. The application of the detection criterion $\tilde{\lambda}^F_1 > \alpha$ leads to the following detection condition:
\begin{equation}
\label{eqn:cond_detection_KF}
\frac{\sigma_T}{\sigma_M} > \frac{\alpha \sqrt{2} }{M}
\end{equation}
This condition is reported in Tab.\ref{tab:perf}. If we compare it with the one obtained for FB (Eq.\ref{eqn:appendix6_1_detection}) and for the classical DORT method (Eq.\ref{eqn:cond_detection_K0}) , we see that the SSF improves the detection condition by a factor $\sqrt{M}$ compared to the DORT mùethod and hence, reaches the level of performance of FB. Actually, it is even slightly better than FB since the threshold $\alpha$ is inferior for DORT applied to $\mathbf{K^F}$. This approach succeeds in detecting the target for higher noise-to-signal ratios, compared to FB. This better performance is reinforced by the robustness of the SSF to aberration (see Sec.\ref{sec:aberration}), which has been neglected in this Appendix. 

\section{\label{sec:app_aberration}Effect of the SSF on aberration}

This appendix deals with the effect of the SSF on aberration. More particularly, we want to express the coefficients $k_{lm}^F$ of the filtered matrix $\mathbf{K^F}$, when aberration exists. The coefficients of matrix $\mathbf{K}$ are given by (Eq.\ref{eqn:eq_target_aberration}):
$$k_{ij}(f) =  d _{ij}(f) \exp \left [ j k \frac{ \left (x_i -X_T \right )^2}{2R_T} \right ] \exp \left [ j k \frac{ \left (x_j -X_T \right )^2}{2R_T} \right ]$$
The coefficients $d_{ij}$ form the distortion matrix $\mathbf{D}$. The first step of the SSF consists in the rotation of data described in Sec.\ref{subsec:rot_anti}. It results in two antidiagonal matrices $\mathbf{A_1}$ et $\mathbf{A_2}$ (Eqs.\ref{eqn:construction_A1} \& \ref{eqn:construction_A2}). Let $\mathbf{A^D_1}$ et $\mathbf{A^D_2}$ be the two antidiagonal matrices built from matrix $\mathbf{D}$ according the same process of Sec. \ref{subsec:rot_anti}:
\begin{eqnarray}
\label{eqn:construction_AD1}
\mathbf{A^D_1}   =  \left [ a^D_{1uv} \right ]  \, & \mbox{of dimension }&(2M-1) \times (2M-1) \mbox{,}\nonumber \\
& \mbox{such that } & a_1^D[u, v]  = d[u + v -1, v - u + 2M -1] \\
\mathbf{A^D_2} =  \left [ a^D_{2uv} \right ]  \, & \mbox{of dimension }&(2M-2) \times (2M-2)\mbox{,}\nonumber \\
& \mbox{such that } & a^D_2[u, v]  = d[u + v , v - u + 2M -1]
\label{eqn:construction_AD2}
\end{eqnarray}
From now on, we will call indifferently $\mathbf{A}$, the matrices $\mathbf{A_1}$ and $\mathbf{A_2}$, and $\mathbf{A^D}$, the matrices $\mathbf{A^D_1}$ and $\mathbf{A^D_2}$.

The coefficients of $\mathbf{A}$ can be expressed with the coefficients of $\mathbf{A^D}$:
\begin{equation}
a_{uv}  =  a^D_{uv} \exp \left [ j k \frac{ y_u^2}{2R_T} \right ] \exp \left [ j k \frac{ \left (y_v- \sqrt{2} X_T \right )^2}{2R_T} \right ] 
\label{eqn:expression_AD}
\end{equation}
with
$$y_u =\frac{x_i - x_j}{\sqrt{2}}\; \mbox {  et  } \; y_v =\frac{x_i + x_j}{\sqrt{2}}$$

The next step of the filter consists in projecting the columns of $\mathbf{A}$ along the characteristic space of single scattering, generated by the vector $\mathbf{S}$ whose coordinates are:
$$s_u = \exp \left [ j k \frac{  y_u^2}{2R_T} \right ]L^{-1/2} $$
The coordinates of the vector $\mathbf{P}$, result of this projection (Eq.\ref{eqn:vector_P}), can be expressed as :
\begin{eqnarray*}
p_v&=&\sum_{u=1}^L s_u^*a_{uv} \\
&=& \frac{1}{\sqrt{L}}\sum_{u=1}^L  \exp \left [- j k \frac{  y_u^2}{2R_T} \right ] a^D_{uv} \exp \left [ j k \frac{y_u^2}{2R_T} \right ] \exp \left [ j k \frac{ \left (y_v - \sqrt{2} X_T \right )^2}{2R_T} \right ] \\ 
&=& \left [ \frac{1}{\sqrt{L}}\sum_{u=1}^L  a^D_{uv} \right ]\exp \left [ j k \frac{ \left (y_v- \sqrt{2} X_T \right )^2}{2R_T} \right ] 
\end{eqnarray*}
The filtered matrix $\mathbf{A^F}$ is finally obtained by multiplying the column vector $\mathbf{S}$ with the line vector $\mathbf{P}$ (Eq.\ref{eqn:Af_coord}). As a result, the coefficients $a^F_{uv}$ are given by:
\begin{eqnarray}
a^F_{uv}&=& s_u p_v\nonumber \\
\label{eqn:expression_AF}
a^F_{uv} &=& \left [ \frac{1}{L}\sum_{u=1}^L  a^D_{uv} \right ] \exp \left [ j k \frac{  y_u^2}{2R_T} \right ] \exp \left [ j k \frac{ \left (y_v- \sqrt{2} X_T \right )^2}{2R_T} \right ]  
\end{eqnarray}
If we compare the expressions of $a_{uv}$ (Eq.\ref{eqn:expression_AD}) and $a^F_{uv}$ (Eq.\ref{eqn:expression_AF}), we see that the SSF averages the distortion coefficients along each column of $\mathbf{A}$. This average corresponds to the term $\left [ \frac{1}{L}\sum_{u=1}^L  a^D_{uv} \right ] $ in Eq.\ref{eqn:expression_AF}. 

Afterwards, two matrices $\mathbf{A_1^F}$ and $\mathbf{A_2^F}$ (containing the filtered antidiagonals) are obtained. The last step consists in constructing the filtered matrix $\mathbf{K^F}$, as described in Sec.\ref{subsec:buiding_Kf} :
\begin{eqnarray}
 & & \mbox{if  }  (l-m)/2  \, \,\mbox{is an integer} \nonumber\\
 & & \mbox{then,  } k^F[l, m]= a_1^F \left [(l - m)/ 2 + M,(l + m)/ 2 \right] \\
& & \mbox{if  }  (l-m)/2  \,\,\mbox{is not an integer} \nonumber \\
 & &\mbox{then,  } k^F[l, m]= a_2^F \left [(l - m -1)/ 2 + M, (l + m -1)/ 2 \right]
\end{eqnarray}
By injecting Eq.\ref{eqn:expression_AF} into the two last equations and reversing the change of coordinates, we obtain:
\begin{equation}
k^F[l, m]= e_{l+m-1} \exp \left [ j k \frac{ \left ( x_l- X_T \right)^2}{2R_T} \right ] \exp \left [ j k \frac{ \left ( x_m- X_T \right)^2}{2R_T} \right ]
\end{equation}
where the coefficients $e_{l+m-1}$ are defined as:
\begin{eqnarray}
 & &\mbox{if  }  (l-m)/2  \, \,\mbox{is an integer,} \nonumber\\
 & & \mbox{then,  }e_{l+m-1}=  \frac{1}{2M-1}\sum_{u=1}^{2M-1}  a_1^D \left [ u, (l + m)/ 2 \right]  \\
 & & \mbox{if  }   (l-m)/2  \,\,\mbox{is not an integer,} \nonumber \\
 & & \mbox{then,  }  e_{l+m-1}=  \frac{1}{2M-2} \sum_{u=1}^{2M-2}  a_2^D \left [u, (l + m-1)/ 2 \right]
\end{eqnarray}
We can finally express coefficients $e_v$ as a function of distortion coefficients $d_{ij}$, using Eqs.\ref{eqn:construction_AD1} \& \ref{eqn:construction_AD2} :
\begin{eqnarray}
 & & \mbox{if  } v \, \,\mbox{is an odd number,} \nonumber\\
 & & \mbox{then,  } e_{v}=  \frac{1}{2M-1}\sum_{u=1}^{2M-1}  d \left [u+\frac{v-1}{2}, \frac{v-1}{2}-u+2M-1 \right ]  \\
 & & \mbox{if  }   v  \,\,\mbox{is an even number,} \nonumber \\
 & & \mbox{then,  } e_{v}=  \frac{1}{2M-2} \sum_{u=1}^{2M-2}  d \left [u+\frac{v}{2},\frac{v}{2}-u+2M-1 \right ] 
\end{eqnarray}
Hence, the coefficients $e_v$ correspond to the average of distortion coefficients along each antidiagonal of the matrix $\mathbf{D}$. The latter equations can be rewritten as:
\begin{eqnarray}
 & & \mbox{if  } v \, \,\mbox{is an odd number,} \nonumber\\
  & & \mbox{then,  }e_{v}=  \left < d \left [u+\frac{v-1}{2}, \frac{v-1}{2}-u+2M-1 \right ] \right >_{u=1,...,2M-1} \\
 & & \mbox{if  }   v  \,\,\mbox{is an even number,} \nonumber \\
& & \mbox{then,  } e_{v}=  \left < d \left [u+\frac{v}{2},\frac{v}{2}-u+2M-1 \right ] \right >_{u=1,...,2M-2}
\end{eqnarray}
where the symbol $<.>$ denotes an average over the variable in the subsript.

\section{\label{sec:app_aberration_2}Detection condition in presence of aberration}

In Sec.\ref{sec:aberration}, we have already shown that the filtered $\mathbf{K^F}$ can de decomposed as follows:
$$\mathbf{K^F} =  \underbrace { \left< \mathbf{K^F} \right > } _{\mbox{Matrix of rank 1}}+ \underbrace {\mathbf{\Delta K^F}}_{\mbox{Random Hankel matrix}}$$ 

$\left< \mathbf{K^F} \right >$ is the mean of $\mathbf{K^F}$, its coefficients are given by:
$$\left< k^{F}_{lm}(f)\right >= \left < e_{l+m-1}(f) \right > \exp \left [ j k \frac{ \left (x_l -X_T \right )^2}{2R_T} \right ] \exp \left [ j k \frac{ \left (x_m -X_T \right )^2}{2R_T} \right ]\mbox{.}$$
The norm of its entries is uniform and equal to the mean of the distortion coefficients, $\left < e_{p} \right >$. The matrix $\left< \mathbf{K^F} \right >$ is of rank 1.

$\mathbf{\Delta K^F}$ corresponds to a perturbation(not necessary small) linked with the fluctuations of distortion coefficients $e_p$. Its coefficients are given by:
$$\delta k^{F}_{lm}(f) = \left [  e_{l+m-1}(f)- \left < e_{l+m-1}(f) \right > \right ]  \exp \left [ j k \frac{ \left (x_l -X_T \right )^2}{2R_T} \right ] \exp \left [ j k \frac{ \left (x_m -X_T \right )^2}{2R_T} \right ]\mbox{.}
$$
The coefficients $\delta k^{F}_{lm}(f)$ are random variables whose standard deviation is $\mbox{std} \left [e_p \right]$. As seen previously, the matrix $\mathbf{\Delta K^F}$ displays the same statistical behavior as a random Hankel matrix.

The aim of this appendix is to determine the ratio $\left | \left < e_{p} \right > \right |/\mbox{std} \left [e_p \right]$, above which the SVD will succeed in extracting the matrix $\left< \mathbf{K^F} \right >$ along the first singular space. To that aim, we will use RMT once again.

We will rely on the same kind of argument as in Appendix \ref{subsec:app2_dort}. Indeed, an analogy can be made with the detection condition found for the signal-to-noise ratio $\sigma_T/\sigma_M$. The signal amplitude $\sigma_T$ corresponds here to the mean value of distortion coefficients $\left | \left < e_{p} \right > \right |$. The standard deviation of $e_p$ plays the role of ``noise'': $\mbox{std}\left [ e_p\right] \Leftrightarrow \sigma_M$ . Note that the matrix $\mathbf{K^F}$ we consider here is of size $\frac{M}{2}\times \frac{M}{2}$. If the first singular space of $\mathbf{K^F}$ corresponds actually to the ``non distorted'' matrix $\left< \mathbf{K^F} \right >$, then the expected value of $\lambda_1^F$ is given by (see Appendix \ref{subsec:app2_dort}):
\begin{equation}
\mbox{E}\left [ \lambda^F_1 \right ]= \frac{M}{2} \left | \left < e_{p} \right > \right |
\end{equation}
The quadratic mean of singular values is given by (see Appendix \ref{subsec:app2_dort}):
\begin{equation}
 \sqrt { \frac{1}{M/2}\sum_{p=1} \left [\lambda^F_p\right ]^2}\simeq\sqrt{\frac{M}{2}\left ( \left | \left < e_{p} \right > \right |^2+  \mbox{var} \left [e_p \right]  \right )}
\end{equation}
Upon normalization (Eq.\ref{eqn:eq2}), the expected value of $\tilde{\lambda}_1^F$ is thus given by:
\begin{equation}
\mbox{E} \left \{ \tilde{\lambda}^F_1 \right \}=\sqrt{\frac{M}{2}\frac{\left | \left < e_{p} \right > \right |^2}{\left | \left < e_{p} \right > \right |^2+\mbox{var} \left [e_p \right] }}\simeq \frac{\left | \left < e_{p} \right > \right |}{ \mbox{std} \left [e_p \right] }\sqrt{\frac{M}{2}} \,\,\mbox{,
for} \,\, \left | \left < e_{p} \right > \right |^2<<\mbox{var} \left [e_p \right] 
\end{equation}
The validity of this approximation $ \left | \left < e_{p} \right > \right |^2<<\mbox{var} \left [e_p \right] $ will be proved \textit{a posteriori} by the final result. As $\mathbf{K^F}$ is a Hankel matrix, the detection criterion is $\tilde{\lambda}^F_1> \alpha$. It leads to the final detection condition :
\begin{equation}
\label{eqn:cond_detection_KF_aberration}
\frac{ \left | \left < e_{p} \right > \right |}{\mbox{std} \left [e_p \right]} > \frac{\alpha}{\sqrt{M/2}}
\end{equation}


\end{document}